\documentclass[aps,prc,twocolumn,showpacs,floatfix,nofootinbib,preprintnumbers,superscriptaddress,amsmath,amssymb]{revtex4-1}

\usepackage{mathrsfs}
\usepackage{graphicx}
\usepackage{epsfig}
\usepackage{bm}
\usepackage{color}
\usepackage{float}
\usepackage{dcolumn}
\usepackage{multirow} 
\usepackage{hyperref}
\usepackage{color}

\graphicspath{{.}}

\newcommand{\be}{\begin{equation}}
\newcommand{\ee}{\end{equation}}
\newcommand{\bea}{\begin{eqnarray}}
\newcommand{\eea}{\end{eqnarray}}

\begin{document}

\title{Effective theory for the non-rigid rotor in an electromagnetic
  field: Toward accurate and precise calculations of
  \textit{E}2 transitions in deformed nuclei}

\author{E. A. Coello P\'erez} 

\affiliation{Department of Physics and Astronomy, University of
  Tennessee, Knoxville, Tennessee 37996, USA}

\author{T. Papenbrock} 

\affiliation{Department of Physics and Astronomy, University of
  Tennessee, Knoxville, Tennessee 37996, USA}

\affiliation{Physics Division, Oak Ridge National Laboratory, Oak
  Ridge, Tennessee 37831, USA}

\date{\today}

\begin{abstract}
  We present a model-independent approach to electric quadrupole
  transitions of deformed nuclei. Based on an effective theory for
  axially symmetric systems, the leading interactions with
  electromagnetic fields enter as minimal couplings to gauge
  potentials, while subleading corrections employ gauge-invariant
  non-minimal couplings. This approach yields transition operators
  that are consistent with the Hamiltonian, and the power counting of
  the effective theory provides us with theoretical uncertainty
  estimates.  We successfully test the effective theory in
  homonuclear molecules that exhibit a large separation of scales.
  For ground-state band transitions of rotational nuclei, the
  effective theory describes data well within theoretical
  uncertainties at leading order.  In order to probe the theory at
  subleading order, data with higher precision would be valuable. For
  transitional nuclei, next-to-leading order calculations and the
  high-precision data are consistent within the theoretical
  uncertainty estimates. We also study the faint inter-band
  transitions within the effective theory and focus on the $E2$
  transitions from the $0^+_2$ band (the ``$\beta$ band'') to the
  ground-state band. Here, the predictions from the effective theory
  are consistent with data for several nuclei, thereby proposing a
  solution to a long-standing challenge.
\end{abstract}


\maketitle

\section{Introduction}
Our understanding of deformed nuclei in the rare-earth and actinide
regions of the nuclear chart is largely based on the geometric
collective models~\cite{bohr1952, bohr1953, eisenberg, bohr1975,
  hess1980, rohozinski2009, rowe2010}, and the algebraic collective
models~\cite{arima1975, matsuo1984}. For even-even nuclei, these
models employ quadrupole degrees of freedom (and an additional $s$
boson in the interacting boson model~\cite{iachello}). The collective
models depend on a small numbers of parameters. They describe the key
features of deformed nuclei, namely low-energy spectra consisting of
rotational bands on top of vibrational band heads, with strong $E2$
intra-band transitions, and much weaker inter-band transitions.
However, some finer details are not well described by the collective
models, and the accurate description of inter-band electromagnetic
transition strengths is a particular challenge. As an example we
mention the overprediction (by factors of 2 to 10) of $E2$ transitions
between the rotational band on top of the $0^+_2$ vibrational band
head (historically called the ``$\beta$ band'') and the ground-state
band for well-deformed nuclei~\cite{garrett2001, aprahamian2004,
  rowe2010}.  This situation is similar for transitional nuclei at the
border between sphericity and deformation.  Here, the models based on
the $X(5)$ solution by \textcite{iachello2001} of the Bohr Hamiltonian
tend to overpredict electromagnetic inter-band
transitions~\cite{zamfir1999,casten2001,kruecken2002,tonev2004}.
  
In recent years, computationally tractable approaches to collective
models~\cite{rowe2004,rowe2009} led to a better understanding of
geometric models and their parameter space~\cite{caprio2005}. However,
it seems that changes to the Bohr Hamiltonian, e.g. by studying
non-separable potentials~\cite{caprio2009} or by considering other
solutions~\cite{inci2011}, do not overcome the deficiencies for the
inter-band transitions.  We also note that a variety of approaches
addressed other shortcomings of the collective models by focusing on
tri-axial deformations~\cite{allmond2008}, or inclusion of isovector
modes~\cite{bentz2011,bentz2014}, see Ref.~\cite{matsuyanagi2010} for
a review of present challenges.
  
Increasing the complexity of collective models, e.g. through the
addition of more terms, can lead to an undesirable proliferation of
parameters and a loss of predictive power. This unattractive feature
of modeling can partly be overcome by effective field theories (EFTs).
An EFT is based on symmetry principles alone and exploits a separation
of scales for the {\it systematic} construction of Hamiltonians based
on a power counting. In this way, an increase in the number of
parameters (i.e. low-energy constants that need to be adjusted to
data) goes hand in hand with an increase in precision, and thereby
counters the loss of predictive power. Furthermore, this systematic
increase in precision makes it possible to estimate theoretical
uncertainties, see \textcite{furnstahl2014c} for a recent review.
Finally, the EFT approach also helps us to identify inherent
limitations that are due to the breakdown scale of the theory.

The successful reproduction of the low-energy spectra of deformed
nuclei strongly suggests that the geometric collective model correctly
captures key aspects such as relevant degrees of freedom and the
interaction between them. This picture is also obtained in a
model-independent approach to deformed nuclei based on
EFT~\cite{papenbrock2011,zhang2013,papenbrock2014}.

The overprediction of the inter-band transition strengths in
collective models thus leads us to scrutinize the operators that are
employed in the calculations of $E2$ transition strengths.  The Bohr
Hamiltonian models the nucleus as an incompressible liquid drop with
quadrupole surface oscillations. These corresponding five degrees of
freedom can be mapped onto three Euler angles (describing overall
rotations of the nucleus) and two deformation parameters (describing
vibrations in the body-fixed coordinate system). In this model, $E2$
transitions are computed from the quadrupole operator.  This approach
to electric transitions in deformed nuclei seems to be motivated by
Siegert's theorem~\cite{siegert1937}, which allows one to employ the
density instead of the current operator in the computation of some
transition rates, see e.g. Ref.~\cite{greiner1997}. We recall that the
derivation of Siegert's theorem is based on gauge invariance and
starts from gauging momentum operators~\cite{sachs1951}. Thus the
applicability of Siegert's theorem is not obvious for the collective
models that employ quadrupole operators for momenta (as opposed
to vectors).

The identification of the transition operator is even more challenging
for the algebraic models because of the lack of a geometric picture.
For the calculation of electromagnetic transition strengths, these
models employ operators that couple the basic degrees of freedom to a
spherical tensor whose rank equals the desired multipole order. For a
recent analysis of this approach, we refer the reader to
Ref.~\cite{ionescu2012}.

In this work we study the electromagnetic coupling of deformed nuclei
within an effective theory motivated by similar approaches
to other nuclear systems, see Refs.~\cite{pastore2009,
  koelling2009,hammer2011, griesshammer2012, hagenp2013,girlanda2014}
for recent examples. In contrast to more phenomenological models, the
consistent treatment of Hamiltonians and currents is a highlight of
effective theories. As we will see, coupling the non-rigid rotor to
electromagnetic fields in a model-independent way is an interesting
problem in itself. Perhaps somewhat surprisingly, we are not aware of
any literature addressing this problem.  Our approach reproduces the
strong intra-band transitions that are also described accurately by
the collective models. For the weaker inter-band transitions, the
effective theory approach yields a much improved description of data
and thereby suggests steps toward overcoming some limitations of the
geometric and algebraic collective models. Finally, the effective
theory approach also permits us to give theoretical uncertainty
estimates and thereby facilitates a meaningful comparison with data.
As we will see, this comparison also suggests that data with higher
precision for $E2$ transitions would be very valuable.

Ultimately, a microscopic theory of deformed nuclei must be based on
fermionic constituents. Nuclear mean field and density functional
theories (see Refs~\cite{nazarewicz1994, frauendorf2001} for reviews),
are making impressive predictions of rotational bands and moments of
inertia~\cite{sheikh2008, niksic2011}, with new projection techniques
being proposed~\cite{duguet2015}. In light $p$-shell nuclei, {\it ab
  initio} approaches are now addressing the emergent behavior of
rotational collective motion~\cite{caprio2013,dytrych2013}. Recently,
fermionic approaches have also been used to constrain parameters of
collective models~\cite{nomura2008}.

This paper is organized as follows. In Sect.~\ref{ET} we briefly
review the effective theory for axially deformed nuclei. The
electromagnetic coupling of the effective theory is described in
Sect.~\ref{EC}. Section~\ref{Intraband} presents the results for
intra-band $E2$ transitions and compares them to data on rotational
and transitional nuclei. A somewhat surprising result is that much of
the available data lacks the precision to challenge the effective
theory.  Sections~\ref{Vibrations} and~\ref{Interband} include
quadrupole degrees of freedom for the description of inter-band
transitions.  Comparison to data shows that the effective theory
accounts well for these faint transitions. Finally, we present our
summary.

\section{Effective theory for the axially symmetric non-rigid rotor}
\label{ET}

In this Section we briefly review the effective theory for deformed
nuclei~\cite{papenbrock2011,zhang2013,papenbrock2014}. The
presentation in this paper aims at being more intuitive and less
technical, though. We first focus on the lowest-energy phenomena and thus on the
axially symmetric non-rigid rotor. The coupling to vibrations is
considered in Sect.~\ref{Vibrations}.

\subsection{Low-energy degrees of freedom}
The effective theory is based on the emergent symmetry breaking from
the rotational symmetry of the group ${\cal G}=\rm{SO(3)}$ to axial
symmetry of the subgroup ${\cal H}=\rm{SO(2)}$. Thus, the Nambu-Goldstone
modes parameterize the
coset~\cite{coleman1969,callan1969,leutwyler1994,roman1999,hofmann1999,kampfer2005,brauner2010}
${\cal G}/{\cal H}=\rm{SO(3)/SO(2)}$ which is isomorph to the
two-sphere.  This agrees with our intuition: the orientation of an
axially symmetric object is defined by two Euler angles or,
equivalently, by the direction of its symmetry axis. In a finite
system, the symmetry breaking has an emergent character, and
(quantized) zero modes take the place of Nambu-Goldstone
modes~\cite{leutwyler1987,hasenfratz1993,papenbrock2014}. In our case,
the polar and azimuthal angles $\theta$ and $\phi$ (also labeled
compactly as $\Omega$) parameterize the two-sphere, i.e., the radial
unit vector
\begin{equation}
\mathbf{e}_{r}\equiv\left(
\begin{array}{c}
\sin{\theta}\cos{\phi}\\
\sin{\theta}\sin{\phi}\\
\cos{\theta}
\end{array}\right)
\end{equation}
indicates the direction of the symmetry axis of the non-rigid rotor.
Thus, the effective theory for this system is
equivalent to that of a particle on the two-sphere.

The velocity of the orientation vector $\mathbf{e}_{r}$ is the time derivative
\begin{equation}
\begin{split}
d_{t}\mathbf{e}_{r} &= \dot{\theta}\mathbf{e}_{\theta}+\dot{\phi}\sin{\theta}\mathbf{e}_{\phi}\\
&\equiv v_{\theta}\mathbf{e}_{\theta}+v_{\phi}\mathbf{e}_{\phi}\equiv \mathbf{v}.
\end{split}
\end{equation}
This vector lies in the plane tangent to the two-sphere at $\Omega$. 
Here and in what follows, we employ dots to denote time derivatives. The
low-energy Lagrangian is a scalar function of the velocity vector
alone and does not depend on the vector $\mathbf{e}_r$ because of the
emergent symmetry breaking. To make progress, we need to understand
the behavior of $\mathbf{v}$ under rotations, and establish a
power counting. With these two ingredients in hand, we then construct
the most general Lagrangian that is consistent with rotational
symmetry and at a given order of the power counting.

\subsection{Rotational invariance}
Under a rotation $r\equiv
r(\alpha,\beta,\gamma)=\exp{(-i\alpha\hat{J}_z)}\exp{(-i\beta\hat{J}_y)}\exp{(-i\gamma\hat{J}_z)}$,
parameterized by the three Euler angles $(\alpha,\beta,\gamma)$, the
angles $\theta$ and $\phi$ transform non-linearly into $\theta'$ and
$\phi'$. This constitutes a nonlinear realization of SO(3).  It is
interesting to comapre this with Bohr's approach to deformed
nuclei~\cite{bohr1952}. Bohr starts from a linear representation of
SO(3) by choosing deformation parameters as coefficients of spherical
harmonics. The transformation to the body-fixed coordinate system then
introduces a nonlinear realization of SO(3) in terms of three rotation
angles and two deformation parameters. The rotation $r$ transforms the
velocity vector $\mathbf{v}(\Omega)$ (or any vector in the tangent
plane) into the vector $\mathbf{v}'(\Omega')$ that lies in the tangent
plane at $\Omega'$. It is clear that the mapping from $\mathbf{v}$ to
$\mathbf{v}'$ is equivalent to a SO(2) rotation in the tangent plane
by an angle $\chi=\chi(\alpha,\beta,\gamma;\Omega)$ that is a
complicated function of the Euler angles and the original coordinates
$\Omega$. Details are given in Ref.~\cite{papenbrock2011}.

At this point it is useful to introduce spherical components of the
velocity inside the tangent plane as
\begin{equation}
v_{\pm}\equiv {1\over\sqrt{2}}(v_{\theta}\pm iv_{\phi}) \ .
\end{equation}
Under a rotation by the Euler angles $(\alpha,\beta,\gamma)$ the
vector $\mathbf{v}$ transforms as
\begin{equation}
  v_{\pm}\rightarrow e^{\mp i\chi}v_{\pm}.
\end{equation}
Thus, under an SO(3) transformation, vectors in the tangent plane
formally transform under an SO(2) transformation, and any Lagrangian
build from elements in the tangent plane that is formally invariant
under SO(2), is in fact invariant under SO(3).

For the general construction of invariant Lagrangians, we must also
consider time derivatives of vectors in the tangent plane. The
resulting vectors may not lie in the tangent plane. Thus, the
ordinary time derivative needs to be replaced by the covariant
derivative
\begin{equation}
\label{Dt}
D_{t}\equiv d_{t}-i\dot{\phi}\cos{\theta}\hat{J}_{z},
\end{equation}
which is the projection onto the tangent plane of the ordinary time
derivative.

Let $L$ denote a rotationally invariant Lagrangian in the
velocities $v_\pm$. The application of Noether's theorem yields the
angular momentum $\mathbf{I}$ as the conserved
quantity~\cite{papenbrock2011}. Its spherical components $I_{+}$,
$I_{0}$ and $I_{-}$ are
\begin{equation}
\begin{split}
\label{spin}
I_{+1} &= -{1\over\sqrt{2}}e^{i\phi}(ip_{\theta}-p_{\phi}\cot{\theta})\\
I_{0} &= p_{\phi}\\
I_{-1} &= -{1\over\sqrt{2}}e^{-i\phi}(ip_{\theta}+p_{\phi}\cot{\theta}).
\end{split}
\end{equation}
Here
\begin{equation}
\label{momenta}
p_{\theta}\equiv\partial_{\dot{\theta}}L\qquad p_{\phi}\equiv\partial_{\dot{\phi}}L
\end{equation}
denotes the canonical momenta. The squared angular momentum is
\begin{equation}
\mathbf{I}^{2}=p_{\theta}^{2}+\frac{p_{\phi}^{2}}{\sin^{2}{\theta}}.
\end{equation}

This construction of the Lagrangian is particularly useful when
further degrees of freedom are coupled to the axially symmetric
rotor.

\subsection{Power counting and the rotational Hamiltonian}
The leading-order (LO) rotationally invariant Lagrangian 
\begin{equation}
\label{rotlo}
L_{\rm LO}=C_{0}v_{+1}v_{-1}=\frac{C_{0}}{2}(\dot{\theta}^{2}+\dot{\phi}^{2}\sin^{2}{\theta})
\end{equation}
is quadratic in the velocities $v_\pm$. It is equivalent to that of a
particle restricted to move on the two-sphere, or to that of a rigid
rotor. Here, $C_0$ is a low-energy constant and corresponds to the
effective moment of inertia. This parameter of our theory must be
fixed by data.

For the power counting, we need to introduce relevant energy scales.
Let $\xi$ denote the low-energy scale associated with rotations. Then,
$\xi\sim 80$~keV and $\xi\sim 40$~keV for deformed rare-earth nuclei and
actinides respectively. The breakdown scale $\omega$ of the effective
theory coincides with the onset of vibrational excitations and is
of the order of 1~MeV and 0.6~MeV for rare-earth
nuclei and actinides respectively. Thus, $\xi/\omega\approx 1/10$ is a
conservative estimate.

The LO Lagrangian and the time derivatives (such as the velocities
$v_\pm$) are of order $\xi$. Thus,
\begin{equation}
\label{power}
v_{\pm}\sim\dot{\phi}\sim\dot{\theta}\sim\xi\qquad C_{0}\sim\xi^{-1}.
\end{equation}

A Legendre transformation of the LO Lagrangian yields the LO
Hamiltonian
\begin{equation}
\label{rothamlo}
H_{\rm LO} = \frac{1}{2C_{0}}\left(p_{\theta}^{2}+\frac{p_{\phi}^{2}}{\sin^{2}{\theta}}\right)= \frac{1}{2C_{0}}\mathbf{I}^{2}.
\end{equation}
The quantization is standard, and the angular momentum $\mathbf{I}$
becomes the angular momentum operator $\hat{\mathbf{I}}$ with
spherical components~\cite{varshalovich1988}
\begin{equation}
\begin{split}
\label{spin_qm}
\hat{I}_{+1} &= -{1\over\sqrt{2}}e^{i\phi}\left(\partial_{\theta}+i\cot{\theta}\partial_\phi\right)\\
\hat{I}_{0} &= -i\partial_{\phi}\\
\hat{I}_{-1} &= -{1\over\sqrt{2}}e^{-i\phi}\left(\partial_{\theta}-i\cot{\theta}\partial_\phi\right).
\end{split}
\end{equation}
The squared angular momentum is
\begin{equation}
\hat{\mathbf{I}}^2=\hat{I}_0^2-\hat{I}_{+}\hat{I}_{-}-\hat{I}_{-}\hat{I}_{+}.
\end{equation}
We also recall that 
\be
\label{defI}
\hat{\mathbf{I}}=\mathbf{e}_r\times \left(-i\mathbf{\nabla}_\Omega\right) \ ,
\ee
with
\be
\mathbf{\nabla}_{\Omega}=\mathbf{e}_{\theta}\partial_{\theta}+\mathbf{e}_\phi{1\over\sin\theta}\partial_\phi
\ee
being the angular derivative in the tangent
plane~\cite{varshalovich1988}. We note that $-i\mathbf{\nabla}_\Omega$
is not an Hermitian operator.

The eigenfunctions of the Hamiltonian~(\ref{rothamlo}) are spherical
harmonics $Y_{IM}(\Omega)$ with eigenvalues 
\be
\label{LOspectrum}
\hat{H}_{\rm LO} Y_{IM}(\theta,\phi) = {I(I+1)\over 2C_0} Y_{IM}(\theta,\phi)
\ee

Higher-order corrections to the LO Lagrangian~(\ref{rotlo}) include
terms with higher powers of $\mathbf{I}^2$.  At next-to-leading order
(NLO) the Lagrangian becomes $L_{\rm LO}+L_{\rm NLO}$ with 
\be
L_{\rm NLO} = {C_2\over 4}\left(\mathbf{I}^2\right)^2. 
\ee
Thus, the corresponding Hamiltonian is $H_{\rm LO}+H_{\rm NLO}$ with
\be 
\label{hamNLO}
H_{\rm NLO}= - {C_2 \over C_0^2} \left(H_{\rm LO}\right)^2 = - {C_2\over 4C_0^3}\left(\mathbf{I}^2\right)^2 \ ,  
\ee 
and the spectrum becomes
\be
\label{NLOspectrum}
E(I)={I(I+1)\over 2C_0} -{C_2\over 4C_0^4} \left(I(I+1)\right)^2 \ .
\ee
This deviation from the rigid-rotor behavior is due to omitted physics
at the energy scale $\omega$ of vibrations. From the expression for
the NLO Hamiltonian (\ref{hamNLO}) it is clear that $C_{2}$ has units
of energy$^{-3}$. The scaling is~\cite{papenbrock2011} 
\begin{equation}\label{NLOpc}
C_{2}\sim C_{0}/\omega^{2},
\end{equation}
and consequently, the ratio of the NLO correction to the LO
contribution of the energy scales as
\begin{equation}\label{NLOtoLO}
\frac{\langle \hat{H}_{\rm NLO}\rangle}{\langle\hat{H}_{\rm LO}\rangle}\sim\left(\frac{\xi}{\omega}\right)^{2}I(I+1).
\end{equation}
Thus, the effective theory of the axially symmetric non-rigid rotor is
identical to the variable-moment-of-inertia
model~\cite{mariscotti1969,scharff1976}, and the spectrum consists of
increasing powers of $I(I+1)$. It is important to notice that
according to Eq.~(\ref{NLOtoLO}), the effective theory is expected to
break down at spins of magnitude $\omega/\xi$, i.e.  when the
second term in Eq.~(\ref{NLOspectrum}) becomes as large as the first
term. For a given nucleus, an estimate for the breakdown spin can be
obtained by employing the LECs $C_0$ and $C_2$.  The result is the
estimate $\sqrt{C_0^3/C_2}$. For the rotors listed in
Table~\ref{LECs}, this estimate usually excceds the general
estimate $\omega/\xi$.

Table~\ref{LECs} below shows values $C_0\xi$, $(C_2/C_0)\omega^2$,
$(\xi/\omega)^2$, and $C_2/C_0^3$ from the description of the
ground-state bands of the homonuclear molecules $N_2$ and $H_2$, the
rotational nuclei $^{236}$U, $^{174}$Yb, $^{166,168}$Er, and
$^{162}$Dy, and the transitional nuclei $^{188}$Os, $^{154}$Gd,
$^{152}$Sm, and $^{150}$Nd, respectively. Here, $\xi$ is the
excitation energy of the lowest $2^+$ state and $\omega$ is the
excitation energy of the lowest vibrational state. The values of $C_0$
and $C_2$ are obtained from a simultaneous fit to the lowest $2^+$ and
$4^+$ levels to Eq.~(\ref{NLOspectrum}), respectively. For a rigid
rotor, $C_0\xi=3$, $\xi/\omega=0$, and $C_2/C_0^3=0$.
Table~\ref{LECs} shows that the ratios $(C_2/C_0)\omega^2$ are of
natural size, i.e. of order one, for the considered molecules and
nuclei, and that the ratios $C_2/C_0^3$ are consistent with (but
systematically smaller than) the scaling estimate $(\xi/\omega)^2$.
This suggests that the breakdown scale is higher than the conservative 
estimate of $\omega$. Still, the values for the LEC $C_2$ are consistent with scaling
estimates. Clearly the molecule N$_2$ is very close to the rigid-rotor
limit.  The comparison suggests that the molecule H$_2$ is as
non-rigid a rotor as the nuclei $^{236}$U, $^{174}$Yb and
$^{168}$Er. The transitional nuclei $^{188}$Os, $^{154}$Gd,
$^{152}$Sm, and $^{150}$Nd exhibit even larger deviations from the
rigid-rotor limit.

\begin{table}[th]
  \begin{ruledtabular}
    \begin{tabular}{cddD{.}{.}{6}D{.}{.}{6}D{.}{.}{6}}
      System & \multicolumn{1}{r}{$C_{0}\xi$} & \multicolumn{1}{r}{${C_{2}\over C_{0}}\omega^2$} & \multicolumn{1}{r}{$(\xi/\omega)^{2}$} & \multicolumn{1}{r}{$C_{2}/C_{0}^{3}$} & \multicolumn{1}{r}{$b/a$} \\\hline
      N$_{2}$    & 3.00 & 2.1 & 0.000026 & 0.000006& -0.000011\\
      H$_{2}$    & 2.99 & 2.2 & 0.0062 & 0.0015 & 0.0022\\
      $^{236}$U  & 2.99 & 2.3 & 0.0043 & 0.0011 &  -\\
      $^{174}$Yb & 2.99 & 3.4 & 0.0026 & 0.0010 & - \\
      $^{168}$Er & 2.99 & 1.0 & 0.0094 & 0.0010 &  -\\
      $^{166}$Er & 2.98 & 1.6 & 0.011 & 0.0020 &  -\\
      $^{162}$Dy & 2.98 & 1.9 & 0.0083 & 0.0017 &  -\\
      $^{154}$Sm& 2.97 & 5.2  & 0.0056 & 0.0033 & - \\ 
      $^{188}$Os & 2.91 & 1.5 & 0.06 & 0.012 & 0.008\\
      $^{154}$Gd & 2.88 & 3.3 & 0.033 & 0.013 & 0.006\\
      $^{152}$Sm & 2.88 & 3.5 & 0.032 & 0.013 & 0.003\\
      $^{150}$Nd & 2.85 & 3.6 & 0.037 & 0.017 & 0.011 \\
    \end{tabular}
  \end{ruledtabular}
  \caption{Dimensionless ratios of LECs and energy scales. The LECs
    $C_0$ and $C_2$ are obtained from the $2^+$ and $4^+$ levels of
    ground-state band for molecules and nuclei considered in this
    work. The ratio $\xi/\omega$ measures the energy scales of
    rotations and vibrations. For a rigid rotor $C_0\xi=3$,
    $\xi/\omega=0$, and $C_2/C_0^3=0$. The ratio $b/a$ measures
    subleading corrections to transition quadrupole moments and
    is similar in size as the subleading energy correction
    $C_2/C_0^3$. A dash indicates that the experimental data is not
    precise enough to determine subleading corrections.}
  \label{LECs}
\end{table}

Within an effective field theory for emergent symmetry breaking in
finite systems~\cite{papenbrock2014}, vibrations enter as the quantized
Nambu-Goldstone modes. The inclusion of vibrations into the theory
pushes the breakdown scale $\Lambda$ to higher energies. We have to
distinguish two cases.  In the first case, $\Lambda$ is set by the
appearance of new degrees of freedom. In nuclei, these are pairing
effects, and $\Lambda\approx 2$~to~3~MeV. In molecules these are
electronic excitations. The second case concerns the breakdown of the
effective theory due to a restoration of spherical symmetry at large
excitation energies. Indeed, for energies $\Lambda\sim \omega^2/\xi$,
the amplitude of vibrations approaches the scale of the static
deformation $\sim \xi^{-1/2}$. In nuclei $\omega^2/\xi\approx
5$~to~10~MeV, and the breakdown scale is thus given by the onset of
new degrees of freedom.

\section{Coupling to electromagnetic fields}
\label{EC}
In this Section, we couple the axially-symmetric non-rigid rotor to
electromagnetic fields. In leading order, minimal couplings of the
gauge fields describe the electromagnetic interaction, and non-minimal
couplings enter as subleading corrections. For the long-wavelength
$E2$ transitions we are interested in, our approach is more technical
than, and differs from, the usual approach taken for the collective
models.  The usual approach is motivated by the result of Siegert's
theorem, that allows one to employ density operators instead of
current operators in transition matrix elements, see
\textcite{eisenberg2} for example. While it is not obvious how to
derive this result for the quadrupole degrees of freedom of the
collective models, Siegert's theorem is expected to hold in leading
order, i.e. for the strong intra-band transitions.  We recall that
\textcite{mikhailov1964,mikhailov1966} employed the quadrupole
operator in the computation of the electromagnetic transition
strengths, and the resulting formulas are well known and widely
used~\cite{bohr1975}. However, this approach fails to describe the
order of magnitude for the faint inter-band transitions.
 
Thus, it is interesting to more formally develop the electromagnetic
theory of the rotor. Within an effective theory one consistently
relates currents to the underlying Hamiltonian.  We also note that
Siegert's theorem does not apply to magnetic
transitions~\cite{austern1951}. The importance of $M1$ transitions is
another motivation for carrying out the formal development.

Deriving the electromagnetic couplings for non-relativistic many-body
systems from first principles is no easy task~\cite{froehlich1993},
see also \textcite{kampfer2005} for a related study within effective
field theory. Here, we follow a simpler path (at the possible cost of
additional LECs). Within an EFT one writes down all gauge-invariant
couplings that are consistent with the underlying symmetries
(rotations, time reversal, and parity), and develops a power counting,
see \cite{pastore2009, koelling2009,hammer2011, griesshammer2012,
  hagenp2013,girlanda2014} for recent examples.  This introduces
minimal couplings (or minimal substitution) and non-minimal couplings.

Before we follow this formal path, however, we briefly consider a
simple three-dimensional system that reduces to the effective theory
under consideration if a ``radial'' degree of freedom is frozen (or
integrated out). This will give us insights into how to gauge the
collective degrees of freedom we are dealing with.  Throughout this
Section we work in the Coulomb gauge and set the scalar electric
potential to zero.

\subsection{Instructive example}

Let us consider a particle of charge $q$ and mass $m$ in a
spherically-symmetric potential $V(r)$ that effectively confines the
particle to a region of thickness $\rho\ll R$ around $r\approx R$. The
Hamiltonian is
\be
\label{simple}
\hat{H}=-{\hbar^2\over 2m}\Delta +V(r),  
\ee
with eigenfunctions $\psi(r,\theta,\phi)=\langle
r\theta\phi|NIM\rangle=[u_N(r)/r]Y_{IM}(\theta,\phi)$. The rotational
excitations are of order $\hbar^2l(l+1)/(2mR^2)$, and much smaller
than radial excitations, which are of order $\hbar^2/(2m\rho^2)$.
Thus, the low-energy spectrum are rotational bands on top of band
heads from radial excitations, and the effective theory developed in 
the previous Section applies. In what follows, we couple
electromagnetic fields to the Hamiltonian~(\ref{simple}). For
transitions within the ground-state band, we can neglect radial
excitations and thereby gain insights into the couplings of a
low-energy effective theory.

We minimally couple $-i\hbar\mathbf{\nabla} \to-i\hbar\mathbf{\nabla}
- q\mathbf{A}$, and keep only the term
linear in $\mathbf{A}$. Thus, the interaction Hamiltonian between the
electromagnetic field and the particle becomes
\bea
\hat{H}^{(\mathbf{A})}&=&i{\hbar q\over 2m} \left(\mathbf{A}\cdot\mathbf{\nabla} +\mathbf{\nabla}\cdot \mathbf{A} \right)
\nonumber\\
&=&i{\hbar q\over 2m} \left(\mathbf{A}\cdot{1\over r}\mathbf{\nabla}_\Omega +\mathbf{A}\cdot\mathbf{e}_r\partial_r\right)\nonumber\\
&&+i{\hbar q\over 2m} \left({1\over r}\mathbf{\nabla}_\Omega\cdot\mathbf{A} +\mathbf{e}_r\partial_r\cdot\mathbf{A}\right) \ .
\eea
We are interested in the long-wavelength limit and assume that the wave 
length $\lambda$ of the electromagnetic field
fulfills $\rho/\lambda\ll 1$. (Note that the systems we are
interested in actually fulfill $R/\lambda\ll 1$.) Thus, the radial
variation of $\mathbf{A}$ can be neglected and we can simply evaluate
this field at $r=R$.  The matrix element that governs electromagnetic
transitions between the initial state $|i\rangle\equiv|N I_i
M_i\rangle$ and final state $|f\rangle\equiv|N I_f M_f\rangle$ within
the band with radial quantum number $N$ is
\bea 
\label{matele}
\lefteqn{\langle f|\hat{H}^{(\mathbf{A})}|i\rangle = i{\hbar q\over 2m}
\bigg(2\langle I_f M_f |\mathbf{A}\cdot\mathbf{e}_r|I_i M_i\rangle \langle N |\partial_r |N\rangle }\nonumber\\
&&+\langle I_f M_f |(\mathbf{A}\cdot\mathbf{\nabla}_\Omega+\mathbf{\nabla}_\Omega\cdot\mathbf{A})|I_i M_i\rangle \langle N|{1\over r}|N\rangle \bigg) \ . 
\eea
We have
\bea
\langle N|{1\over r}|N\rangle = \int\limits_0^\infty dr {u_N^2(r)\over r}\approx& R^{-1}
\eea
for wave functions that are localized to a small region $\rho\ll R$
around $r\approx R$.  Corrections to this expression are of order
$\rho/R$.

Likewise,
\bea
\langle N |\partial_r |N\rangle &=& \int\limits_0^\infty dr r^2 {u_N(r)\over r} \partial_r {u_N(r)\over r} \nonumber\\
&=&
\int\limits_0^\infty dr \left( u_N(r) u_N'(r) -{u_N^2(r)\over r}\right) \nonumber\\
&\approx& -R^{-1} \ , 
\eea
because the first term vanishes due to $u_N(0)=0=u_N(\infty)$, and the
second term again yields approximately $-1/R$. Again, corrections are
of order $\rho/R$.

Thus, for intra-band transitions the matrix element that governs
long-wavelength transitions becomes in leading order of $\rho/R$
\bea
\label{matele_eft}
\langle f|\hat{H}^{(\mathbf{A})}|i\rangle &\approx& i{\hbar q\over 2m R}
\Big(\langle I_f M_f |(\mathbf{A}\cdot\mathbf{\nabla}_\Omega+\mathbf{\nabla}_\Omega\cdot\mathbf{A})|I_i
M_i\rangle \nonumber\\ &&+ 2\langle I_f M_f
|\mathbf{A}\cdot\mathbf{e}_r|I_i M_i\rangle \Big) \ .  
\eea 
We note that this leading-order expression is independent of the
confining radial potential, and it becomes exact in the limit $\rho/R\to
0$. We also note that the right-hand side of Eq.~(\ref{matele_eft})
does not reference the radial wave function. However, the term
$\mathbf{A}\cdot\mathbf{e}_r$ originates from the current associated
with the radial zero-point motion.  Thus, in a low-energy effective
theory, electromagnetic transitions are induced by the operator
\bea
\label{em_op}
\lefteqn{\hat{H}^{(\mathbf{A})}(\Omega) = -{q\hbar\over 2mR}}\nonumber\\
&&\times\left(\mathbf{A}\cdot \left(-i\mathbf{\nabla}_\Omega +i\mathbf{e}_r \right) 
+\left(-i\mathbf{\nabla}_\Omega +i\mathbf{e}_r \right)\cdot\mathbf{A}\right),
\eea 
and corrections are of order $\rho/R$. We note that the operator  
\be
\label{op}
-i \mathbf{\nabla}_\Omega +i \mathbf{e}_r = {i\over 2}\left[  \hat{\mathbf{I}}^2, \mathbf{e}_r \right] 
\ee
(unlike the operator $-i \mathbf{\nabla}_\Omega$) is also Hermitian
under the usual integration measure ${\rm d}\Omega\equiv{\rm d}\phi
{\rm d}\theta \sin\theta$ of the sphere. The identity~(\ref{op}) can
be proved by a direct computation.

On the first view it might be surprising that the
operator~(\ref{em_op}), relevant for the coupling of the low-energy
degrees of freedom (the angles $\Omega$), references the radial component of the
electromagnetic field $\mathbf{A}$. Indeed, decomposing the vector potential
\bea
\mathbf{A}&=&A_r \mathbf{e}_r +\mathbf{A}_\Omega \\
\mathbf{A}_\Omega &=& A_\theta\mathbf{e}_\theta + A_\phi\mathbf{e}_\phi  
\eea
into a radial component and the projection 
$\mathbf{A}_\Omega$ on the tangential plane, 
and using the identity
\be
-i\mathbf{\nabla}_\Omega \cdot \mathbf{A} = -i\mathbf{\nabla}_\Omega \cdot \mathbf{A}_\Omega -i2\mathbf{e}_r\cdot\mathbf{A}
\ee
we can rewrite the interaction~(\ref{em_op}) as
\bea
\label{op_em3}
\hat{H}^{(\mathbf{A})}(\Omega)
&=&i {q\hbar\over 2mR}\left(\mathbf{A}_\Omega \cdot\mathbf{\nabla}_\Omega +\mathbf{\nabla}_\Omega\cdot\mathbf{A}_\Omega\right) \ .
\eea
This result is in keeping with expectations that a low-energy
effective theory only involves low-energy degrees of freedom.  While
this expression reflects that the physics is entirely in the
tangential plane, it is not ideal because of the appearance of the
non-Hermitian operator $-i\mathbf{\nabla}_\Omega$.  An equivalent
expression involving only Hermitian operators can be obtained using
the angular momentum operator~(\ref{defI}). This yields
\be
\label{op_em4}
\hat{H}^{(\mathbf{A})}(\Omega)
= -{q\hbar\over 2mR}\left[\left(\mathbf{e}_r\times\mathbf{A}_\Omega\right) \cdot \hat{\mathbf{I}} +\hat{\mathbf{I}}\cdot\left(\mathbf{e}_r\times \mathbf{A}_\Omega\right)\right] \ .
\ee
The interaction terms~(\ref{op_em3}) and (\ref{op_em4}) thus suggest that the electromagnetic coupling is achieved by gauging
\be
-i\mathbf{\nabla}_\Omega \rightarrow -i\mathbf{\nabla}_\Omega - q\mathbf{A}_\Omega
\ee
and, equivalently,  
\be
\hat{\mathbf{I}}\rightarrow\hat{\mathbf{I}}-q\mathbf{e}_r\times \mathbf{A}_\Omega \ . 
\ee
The next Subsection confirms this picture. 
 
\subsection{Gauging the effective theory}

Let us now turn to couple electromagnetic fields to the non-rigid
rotor.  The LO effective theory starts from the
Hamiltonian~(\ref{rothamlo}).  Requiring invariance under local gauge
transformations $\psi(\Omega)\to \exp{(i\lambda(\Omega))}\psi(\Omega)$
of its eigenfunctions $\psi(\Omega)$ introduces gauge fields according
to
\begin{equation}
\label{gauging}
\hat{\mathbf{I}}\rightarrow\hat{\mathbf{I}}-q\mathbf{e}_r\times \mathbf{A}_\Omega \ , 
\end{equation}
with 
\be
\mathbf{A}_\Omega=-\mathbf{\nabla}_\Omega\lambda(\Omega) \ .
\ee
Here, the effective charge $q$ is a LEC and needs to be adjusted to data. 
Thus, the requirement of local gauge invariance introduces gauge
fields with components in the tangential plane spanned by the vectors
$\mathbf{e}_\theta$ and $\mathbf{e}_\phi$. As 
$\mathbf{A}_\Omega\cdot\mathbf{e}_r=0$, we 
have $\mathbf{e}_r\times\mathbf{A}_\Omega = \mathbf{e}_r\times\mathbf{A}$,
and this can be employed in the minimal coupling~(\ref{gauging}).  

We are interested in single-photon transitions, and the LO Hamiltonian
that describes the non-rigid rotor plus electromagnetic fields system becomes
\bea
\hat{H}_{\rm LO}^{\rm EM}=\hat{H}_{\rm LO}+\hat{H}^{(\mathbf{A})}_{\rm LO}
\eea
with the interaction Hamiltonian given by
\bea
\label{gauge1}
\hat{H}^{(\mathbf{A})}_{\rm LO}&=&-{q\over 2C_0}\left((\mathbf{e}_r\times\mathbf{A}) \cdot\hat{\mathbf{I}} + \hat{\mathbf{I}} \cdot (\mathbf{e}_r\times\mathbf{A})\right)\nonumber\\
&=&i {q\over 2C_0}\left(\mathbf{A}_\Omega\cdot\mathbf{\nabla}_\Omega + \mathbf{\nabla}_\Omega\cdot\mathbf{A}_\Omega\right) \ .
\eea
This is essentially the operator~(\ref{op_em3}). Thus, the gauging of
the effective theory yields the same interaction Hamiltonian as the
removal of a high-energy degree of freedom in the direct calculation
presented in the previous Subsection. The direct use of the
operator~(\ref{gauge1}) in the computation of matrix elements is
cumbersome. Instead we return to Eq.~(\ref{em_op}), use the
identity~(\ref{op}), and find
\bea
\label{ids}
\left(\mathbf{e}_r\times\mathbf{A}_\Omega\right) \cdot \hat{\mathbf{I}}
&=& 
{i\over 2}\mathbf{A} \cdot\left[
  \hat{\mathbf{I}}^2, \mathbf{e}_r \right] - i \mathbf{A}\cdot\mathbf{e}_r\nonumber\\
\hat{\mathbf{I}}\cdot\left(\mathbf{e}_r\times\mathbf{A}_\Omega\right) &=& 
{i\over 2}\left[
  \hat{\mathbf{I}}^2, \mathbf{e}_r \right]\cdot\mathbf{A} + i \mathbf{A}\cdot\mathbf{e}_r\ .
\eea
Thus, in the long wave length limit and
in LO of the effective theory, the interaction Hamiltonian is
\be
\label{em_op2}
\hat{H}^{(\mathbf{A})}_{\rm LO}=-{i q\over 4C_0}\left(\mathbf{A}\cdot \left[\hat{\mathbf{I}}^2, \mathbf{e}_r \right] 
+\left[\hat{\mathbf{I}}^2, \mathbf{e}_r \right]\cdot\mathbf{A}\right)  \ .  
\ee
This LO interaction Hamiltonian~(\ref{em_op2}) can be rewritten by
employing the LO Hamiltonian~(\ref{rothamlo}) of the rigid rotor,
yielding
\be
\label{em_op3}
\hat{H}^{(\mathbf{A})}_{\rm LO}=-{i q\over 2}\left(\mathbf{A}\cdot \left[\hat{H}_{\rm LO}, \mathbf{e}_r \right] +\left[\hat{H}_{\rm LO}, \mathbf{e}_r \right]\cdot\mathbf{A}\right). 
\ee

At NLO, we start from the Hamiltonian~(\ref{hamNLO}) and minimally
couple it according to Eq.~(\ref{gauging}). Again, we only keep terms
linearly in $\mathbf{A}$ because we are interested in single-photon
transitions. This yields 
\bea
\hat{H}_{\rm NLO}^{\rm EM}=\hat{H}_{\rm NLO}+\hat{H}^{(\mathbf{A})}_{\rm LO}
+\hat{H}^{(\mathbf{A})}_{\rm NLO}\ .
\eea
Here, the NLO interaction takes the form
\bea
\label{ANLO}
\hat{H}^{(\mathbf{A})}_{\rm NLO}&=& 
{q C_2\over 4C_0^4}
\bigg((\mathbf{e}_r\times\mathbf{A})\cdot\hat{\mathbf{I}} +\hat{\mathbf{I}}\cdot (\mathbf{e}_r\times\mathbf{A})\bigg)\hat{\mathbf{I}}^2\nonumber\\
&&+{q C_2\over 4C_0^4}\hat{\mathbf{I}}^2 \bigg((\mathbf{e}_r\times\mathbf{A})\cdot\hat{\mathbf{I}} + \hat{\mathbf{I}}\cdot(\mathbf{e}_r\times\mathbf{A})
\bigg)\nonumber\\
&=& -{C_2\over 2C_0^3}\left(\hat{H}^{(\mathbf{A})}_{\rm LO} \hat{\mathbf{I}}^2 + \hat{\mathbf{I}}^2 \hat{H}^{(\mathbf{A})}_{\rm LO}\right) \ . 
\eea
Note that the LECs of $\hat{H}^{(\mathbf{A})}_{\rm NLO}$ are determined
entirely by the Hamiltonian~(\ref{hamNLO}) and the LO electromagnetic
transitions.  This is the consistency between currents and Hamiltonian
offered within an effective theory. This term is a factor $C_2/C_0^3\sim(\xi/\omega)^2$ smaller than
$\hat{H}^{(\mathbf{A})}_{\rm LO}$. Let
\be
\label{mateleLO}
M_{\rm LO}(i\to f)\equiv \langle f |\hat{H}^{(\mathbf{A})}_{\rm LO}|i\rangle
\ee
be the LO matrix element for electromagnetic transitions. Then
\bea
\lefteqn{M_{\rm NLO}(i\to f)\equiv \langle f |\hat{H}^{(\mathbf{A})}_{\rm NLO}|i\rangle} \nonumber\\
&&= -{C_2\over 2C_0^3}\left[I_f(I_f+1)+I_i(I_i+1)\right]M_{\rm LO} \ .
\eea

We will employ a multipole expansion. This expansion is valid if the
wavelength of the radiation is considerably larger than the linear
dimension of the rotor. Let $k$ be the wave number of the
electromagnetic field.  We have $k\sim\xi$ for transitions in the
ground-state band. For a rigid rotor with extension $R$ and mass $M$,
$C_0\sim MR^2\sim \xi^{-1}$. Thus, $kR\sim\sqrt{\xi/M}$. To give
quantitative estimates, we consider rare earth nuclei. Here,
$kR\approx 1/300$. Thus, the multipole expansion is rapidly
converging.

To make progress, we employ a plane wave
\be
\mathbf{A}(\mathbf{r},t)=A \mathbf{e}_z e^{i(\mathbf{k}\cdot\mathbf{r} -w t)}
\ee
with amplitude $A$, polarization $\mathbf{e}_z$ in the $z$ direction and momentum
$\mathbf{k}=k\mathbf{e}_x$ in the $x$ direction. Here $w=k$ (recall
that the speed of light $c=1$). Taylor expansion of the plane wave yields the leading-oder 
quadrupole component contained in the term
\bea
\mathbf{A}^{(2)}=A\mathbf{e}_z kr \cos\phi \sin\theta \ .
\eea
In what follows, we neglect the subleading contribution of
$\mathbf{A}^{(2)}$ to dipole transitions.  When inserted into the LO
interaction Hamiltonian~(\ref{em_op3}), we find 
\be
\label{E2LO}
H^{(\mathbf{A}^{(2)})}_{\rm LO}=-\frac{i q}{2}\left[H_{\rm LO},
  \mathbf{A}^{(2)}\cdot \mathbf{e}_r \right].  
\ee 
This form of the quadrupole interaction is particularly suited for the
computation of the quadrupole transition matrix
elements~(\ref{mateleLO}), and
\be
\label{mateleLO2}
M_{\rm LO}(E2,i\to f)= -\frac{i qw}{2}\langle f
|\mathbf{A}^{(2)}\cdot\mathbf{e}_r|i\rangle 
\ee 
Here, $w=E_f-E_i$ is the difference between the LO energies of the
final and initial states. The corresponding NLO interaction
Hamiltonian can be obtained directly by inserting the Hamiltonian
(\ref{E2LO}) into Eq.~(\ref{ANLO}). At NLO, the matrix element for
electric quadrupole transitions is equivalent to that of
Eq. (\ref{mateleLO2}), with $w$ being the difference between the NLO
energies of the final and initial states. In the evaluation of these
matrix elements, we will set $r = R$, and absorb the factor $kR$ by
re-defining $qkR\to q$.

\subsection{Non-minimal couplings}

Non-minimal couplings (i.e. interaction terms that include electric
and magnetic fields) arise because the low-energy degrees of freedom
we employ describe composite objects.  Such terms are gauge-invariant
scalars that are consistent with the symmetries of the effective
theory.  For electric transitions, we can couple the low-energy
degrees of freedom to the electric field $\mathbf{E}$, and the power
counting is in derivatives on the electric field and low-energy
degrees of freedom. In leading order we have
\bea
\label{non-minimal-lo}
\hat{H}^{(\mathbf{E})}_{\rm LO} \equiv d \mathbf{E}\cdot\mathbf{e}_r \ .
\eea
Here, the dimensionless number $d$ is a LEC and has to be adjusted to
data. We note that $\mathbf{E}\sim \xi\mathbf{A}$ for low-energy
transitions and assume that $d\sim{\cal O}(1)$. Thus, the non-minimal
term~(\ref{non-minimal-lo}) is of the same order as
$\hat{H}^{(\mathbf{A})}_{\rm LO}$ in Eq.~(\ref{em_op3}).

For the $E2$ transitions considered in this work
$\mathbf{E}^{(2)}=iw\mathbf{A}^{(2)}$, and its is clear that the
transition matrix element of the non-minimal
interaction~(\ref{non-minimal-lo}) is equivalent to the LO gauged
interaction $\hat{H}^{(\mathbf{A})}_{\rm LO}$ after identifying the LECs
$d=q$. We thus see that Siegert's theorem is valid for the LO transitions.

We turn to higher-order non-minimal couplings. In principle, every
single term that is invariant under gauge transformations, rotations,
parity and time reversal must be considered. However, the power
counting~(\ref{power}) establishes which terms are relevant at each
order. The relevant NLO terms are quadratic in $\mathbf{I}$
\begin{equation}\label{nonmin}
\begin{split}
\hat{H}^{(\mathbf{E})}_{\rm NLO}= &- \frac{qd_{1}}{4}\left(\mathbf{E}\cdot\mathbf{e}_{r}\hat{\mathbf{I}}^{2}+\hat{\mathbf{I}}^{2}\mathbf{E}\cdot\mathbf{e}_{r}\right)\\
&- \frac{qd_{2}}{4}\left(\mathbf{E}\cdot\hat{\mathbf{I}}^{2}\mathbf{e}_{r}+\mathbf{e}_{r}\cdot\hat{\mathbf{I}}^{2}\mathbf{E}\right),
\end{split}
\end{equation}
where the factor $q/4$ is included for convenience. As a NLO correction, it is expected to fulfill a relation similar to that of Eq.~(\ref{NLOtoLO})
\begin{equation}\label{errest}
\frac{\langle f|\hat{H}^{(\mathbf{E})}_{\rm NLO}|i\rangle}{\langle f|\hat{H}^{(\mathbf{E})}_{\rm LO}|i\rangle}\sim\left(\frac{\xi}{\omega}\right)^{2}f(I_{i},I_{f}),
\end{equation}
where $f(I_{i},I_{f})$ is a function of the angular momenta of the
initial and final states. From here, it is expected that
$d_{1}\sim d_{2}\sim(\xi/\omega)^{2}$. These LECs need to be
fitted to data.

In this work, we are only interested in electric transitions. For
magnetic transitions, other non-minimally coupled terms involving the
magnetic field $\mathbf{B}$ must be included.

\section{Transitions within the ground band}
\label{Intraband}

In this Section, we study electric transitions within ground-state
bands of molecules and atomic nuclei. Molecules are a perfect testing
ground for the effective theory because the separation of scale
between rotations and vibrations is several orders of magnitude. After
a brief discussion of molecules we consider rotational nuclei in the
rare-earth and actinide regions. For these, the separation of scale
between rotations and vibrations is largest in atomic nuclei. Finally,
we consider transitional nuclei where the separation of scale is
smaller, and NLO corrections are more prominent. A list of rotors
studied in this Section is shown in Table~\ref{tab:alpha}. For a rigid
rotor, $\xi/\omega=0$, and $E_{4^+}/E_{2^+}=10/3$. The other columns
in Table~\ref{tab:alpha} will be discussed below.

\begin{table}[h]
  \begin{tabular}{c|lllll}
    \hline\hline
    Rotor & $\xi/\omega$ & $E_{4^+}/E_{2^+}$ & $Q_{0}$[eb] & $\alpha_{\rm LO}$ & $\alpha_{\rm NLO}$ \\
    \hline
    N$_{2}$    & 0.005& 3.33 & 1.00\footnotemark[1] & 2.18 & 0.70\\
    H$_{2}$    & 0.08 & 3.30 & 1.00\footnotemark[1] & 1.45 & 0.10 \\ 
    $^{236}$U  & 0.05 & 3.30 & 3.29 & 0.00 & -- \\
    $^{174}$Yb & 0.05 & 3.31 & 2.44 & 1.07 & -- \\
    $^{168}$Er & 0.10 & 3.31 & 2.42 & 3.02 & -- \\
    $^{166}$Er & 0.10 & 3.29 & 2.42 & 0.00 & -- \\
    $^{162}$Dy & 0.09 & 3.29 & 2.29 & 0.33 & -- \\
    $^{154}$Sm & 0.07 & 3.25 & 2.08 & 0.23 & -- \\
    $^{188}$Os & 0.24 & 3.08 & 1.58 & 0.32 & 0.43 \\
    $^{154}$Gd & 0.18 & 3.01 & 1.96 & 0.35 & 0.00 \\
    $^{152}$Sm & 0.18 & 3.01 & 1.86 & 0.20 & 0.00 \\
    $^{150}$Nd & 0.19 & 2.93 & 1.65 & 0.38 & 0.32 \\
    \hline\hline
  \end{tabular}
  \footnotetext[1]{Arbitrary units used for molecules.}
  \caption{Ratio $\xi/\omega$ and ratio $E_{4^+}/E_{2^+}$ of energies
    $E_{J^\pi}$ of states with spin $J$ and parity $\pi$ (as measures
    of the separation of scale), and the effective quadrupole moment $Q_{0}$
    for molecules and nuclei considered in this work. For a rigid rotor, 
    $\xi/\omega=0$, and $E_{4^+}/E_{2^+}=10/3$. The constants 
    $\alpha_{\rm LO}$ and $\alpha_{\rm NLO}$ are obtained from  
    $\chi^2$ fits at LO and NLO, respectively, and indicate the size 
    of theoretical errors required to achieve a $\chi^2\approx 1$ per degree of freedom.}
  \label{tab:alpha}
\end{table}

\subsection{Transition strengths}
The reduced transition probabilities of electric radiation with
multipolarity $\lambda$, i.e. the $B(E\lambda)$ values, are given by
Fermi's golden rule
\begin{equation}
\label{be2}
B(E\lambda,i\rightarrow f)=\frac{1}{2l_{i}+1}\left|\langle f||\hat{\mathscr{M}}(E\lambda)||i\rangle\right|^{2},
\end{equation}
where
$\hat{\mathscr{M}}(E\lambda)\equiv(\hat{H}^{(\mathbf{A})}+\hat{H}^{(\mathbf{E})})/wA$.
As we will see below, these transition strengths contain a simple
geometrical factor that governs the leading angular-momentum
dependence. To understand transition strengths within an effective
theory, it is very useful to remove this trivial factor. For this
reason we define the quadrupole transition moments $Q_{if}$ as
\begin{equation}
Q_{if}^{2}\equiv\frac{B(E2,i\rightarrow f)}{\left(C_{I_{i}020}^{I_{f}0}\right)^{2}}. 
\end{equation}
Here $C_{I_{1}M_{1}I_{2}M_{2}}^{I_{3}M_{3}}$ is a Clebsch-Gordan
coefficient~\cite{varshalovich1988} and governs the leading
angular-momentum dependence.

If the quadrupole components of the vector potential $\mathbf{A}$ and
the corresponding electric field $\mathbf{E}$ are inserted into the
transition operators $\hat{H}^{(\mathbf{A})}$ and
$\hat{H}^{(\mathbf{E})}$, they induce $E2$ transitions. At NLO, the
$B(E2)$ values for decays within the ground-state band are
\begin{equation}\label{decaybe2_top}
B(E2,i\rightarrow f)=\frac{(aqR)^{2}}{60}\left(C_{I_{i}020}^{I_{f}0}\right)^{2}\left[1+\frac{b}{a}I_{i}(I_{i}-1)\right] .
\end{equation}
Here $a=1+d_{1}$ and $b=2(d_{1}+d_{2})$ are combinations
of LECs from the non-minimal couplings.
Thus, the quadrupole transition moments for these decays are given by
\begin{equation}\label{decaybe2_res}
Q_{if}^{2}=\frac{(aqR)^{2}}{60}\left[1+\frac{b}{a}I_{i}(I_{i}-1)\right]
\end{equation}
or
\begin{equation}\label{decaybe2}
Q_{if}^{2}=Q_{0}^{2}\left[1+{b\over a}I_{i}(I_{i}-1)\right]
\end{equation}
where $Q_{0}\equiv\sqrt{(aqR)^{2}/60}$ may be thought of as the
effective quadrupole moment. Table~\ref{tab:alpha} shows the values of
$Q_{0}$ for the systems considered in this work. They are obtained
from a global fit to data presented in the second half of this Section. 

In LO the effective theory thus predicts that the quadrupole
transition moments $Q_{if}$ are constant, reflecting the behavior of a
rigid rotor. The NLO corrections are deviations from this behavior
that are quadratic in the angular momentum of the initial state.  They
scale as $d_{1}+d_{2}\sim(\xi/\omega)^{2}$. We note that the NLO
corrections to the quadrupole transitions are thus similar in size and
functional form to the NLO correction of the spectrum of the
ground-state band.

It is interesting to compare the results from the effective theory
with the geometric collective model.  According to
\textcite{bohr1975}, the reduced matrix elements for quadrupole decays
within the ground band are
\bea
\lefteqn{\langle f||\mathscr{M}(E2)||i\rangle=M_{1}(2I_{i}+1)^{1/2}C_{I_{i}020}^{I_{f}0}}\label{bohrmatrixelement}\\
&&\times \left[1+2\frac{M_{2}}{M_{1}}+4\frac{M_{3}}{M_{1}}+2\left(\frac{M_{2}}{M_{1}}+8\frac{M_{3}}{M_{1}}\right)I_{i}(I_{i}-1)\right]. \nonumber
\eea
Here $\mathscr{M}(E2)$ is the quadrupole operator, and $M_{i}$,
$i=1,2,3$ are intrinsic matrix elements. From the matrix
elements~\ref{bohrmatrixelement}, the quadrupole transition moments
for decays within the ground band are
\begin{equation}
Q_{if}^{2}=(a_{\rm BH}M_{1})^{2}\left[1+\frac{b_{\rm BH}}{a_{\rm BH}}I_{i}(I_{i}-1)\right],
\end{equation}
with $a_{\rm BH}=1+2M_{21}+4M_{31}$, $b_{\rm BH}=4(M_{21}+8M_{31})$
and $M_{ij}=M_{i}/M_{j}$. Thus, the effective theory at NLO reproduces 
the geometric collective model and gives the same description for
decays within the ground band. A novel aspect of the effective theory 
is the estimate of theoretical uncertainties.

\subsection{Estimate of theoretical uncertainties}
\label{uncertainties}

The estimate of theoretical uncertainties is a highlight of effective
field theories, see~\textcite{furnstahl2014c} for a recent overview,
and Ref.~\cite{dobaczewski2014} for a general discussion. So far, such
estimates are virtually absent when phenomenological collective models
are applied to describe data.  In effective field theories, the
existence of a breakdown scale and the ensuing power counting allows
one to consistently estimate the size of missing contributions. For
example, when making LO fits to energy levels or quadrupole
transitions in ground-state bands, relative theoretical uncertainties
involving a state with spin $I$ scale as
\be
\varepsilon\equiv(I\xi/\omega)^2 \ .
\ee
At NLO, the relative theoretical uncertainty scales as
$\varepsilon^2$, etc.  The effective theory yields uncertainty
estimates, i.e. it predicts the scale of the theoretical error, but
not its precise absolute size $\alpha\varepsilon^n$.  The expectation
is that $\alpha$ be of natural size, i.e.  $1/3 \lesssim \alpha
\lesssim 3$ or so.  In other words, for a natural value $\alpha$ of
order one, the relative error is of order ${\cal O}(\varepsilon^n)$ at
the $n$~th order in the effective theory. Choosing a natural-size
value for $\alpha$ is thus a simple way to present theoretical
uncertainty estimates, similar to the idea of presenting
order-of-magnitude estimates for remainders in polynomial
approximations to functions.  For consistency, one would expect that
uncertainty estimates for increasing order overlap with each other.

In what follows we will choose $\alpha$ such that a $\chi^2$ per degree of freedom of 1 results from a fit to data. Theoretical
uncertainties can then be viewed as the usual one-sigma bands. One
expects that the resulting value for $\alpha$ is of natural size. A
value of $\alpha\ll 1$ ($\alpha\approx 0$) indicates that the theory
with very small (vanishing) theoretical uncertainties already
describes the data within the experimental error bars.  In such a
case, the the data is not sufficiently precise to challenge the
theory, and we will choose a natural-size value for $\alpha$ for
uncertainty estimates. A very large value $\alpha \gg 1$ signals the
breakdown of the effective theory, because the assumed separation of
scales is not reflected in the data.

The LECs $C_0$ and $C_2$ that govern the spectrum are computed from
the experimental energies of the $2^+$ and $4^+$ states in the
ground-state rotational band. The uncertainty of these LECs can be
neglected because energies are known very precisely.

Let us turn to quadrupole transitions. Here the LECs are $Q_0$ at LO,
and the ratio $b/a$ at NLO.  We denote the (constant) transition
strength at LO as $Q_{\rm LO}$. Its theoretical uncertainty is
\begin{equation}
\sigma_{\rm th}= \alpha_{\rm LO} \frac{C_{2}}{C_{0}^{3}}I_{i}(I_{i}-1) Q_{\rm LO}  \ .
\end{equation}
At NLO, the theoretical uncertainty is given in terms of the NLO
result $Q_{\rm NLO}$ as
\begin{equation}
\sigma_{\rm th}= \alpha_{\rm NLO} \left[\frac{C_{2}}{C_{0}^{3}}I_{i}(I_{i}-1) \right]^2 Q_{\rm NLO} \ . 
\end{equation}
To determine the LECs $Q_0$ and $b/a$ involved in the quadrupole
transitions, we perform $\chi^2$ fits to data, with
\be
\chi^2 = \sum_{d} \frac{\left[Q_{\rm exp}(d) - Q_{\rm th}(d)\right]^2}{\sigma^2_{\rm exp} (d)+\sigma^2_{\rm th}(d)} \ .
\ee
Here, the sum is over all data points, $Q_{\rm exp}(d)$ ($Q_{\rm
  th}(d)$) is the experimental (theoretical) value, and $\sigma_{\rm
  exp}$ the experimental uncertainty. We adjust $\alpha_{\rm LO}$
($\alpha_{\rm NLO}$) in LO (NLO) fits such that the resulting $\chi^2$
per degree of freedom is 1.

Table~\ref{tab:alpha} shows the values of $\alpha_{\rm LO}$ and
$\alpha_{\rm NLO}$ that result from the $\chi^2$ fits. Some of the
fits result in a $\chi^2$ per degree of freedom below 1 even for
vanishing theoretical uncertainty. In such cases, $\alpha_{\rm LO}=0$
or $\alpha_{\rm NLO}=0$. This happens if the theoretical prediction
(with zero theoretical uncertainty estimates) aleady describes all
data within the experimental uncertainties alone. In these cases, we
will employ $\alpha_{\rm LO}=1$ ($\alpha_{\rm NLO}=\alpha_{\rm LO}$)
in LO (NLO) estimates of theoretical uncertainties in the following
Subsections. The values of $\alpha$ in Table~\ref{tab:alpha} are
mostly of natural size. This indicates that the effective theory
describes the data consistently.

Below we will see that experimental uncertainties for quadrupole
transitions are significant and presently preclude us from making any
meaningful subleading predictions for the rotational nuclei $^{236}$U,
$^{174}$Yb, $^{166,168}$Er, $^{162}$Dy, and $^{154}$Sm.  The situation
is better though for the transitional nuclei $^{188}$Os, $^{154}$Gd,
$^{152}$Sm, and $^{150}$Nd, where data with higher relative precision
is available. To test the effective theory for physical systems close
to the rigid-rotor limit, we therefor consider the homonuclear
molecules $H_2$ and $N_2$.

\subsection{Linear molecules}
Linear molecules provide an ideal testing ground for the effective
theory, because they are axially symmetric in their ground states and
close to the rigid rotor limit.  For these molecules, the separation
of scale is excellent, and a good agreement between the effective
theory and experimental data must be achieved at low order.

Homonuclear molecules appear in two isomeric forms, depending on the
alignment of the nuclear spins. For antiparallel spins (the ``para''
state), the system posses a positive $\mathcal{R}$ parity as rotations
of $\pi$ around any axis perpendicular to the symmetry axis do not
change the wave function of the system. This symmetry implies that
only states with even spin $I$ are allowed in the ground band. Thus,
within the ground band, $E2$ transitions are the most relevant, and
this property is shared with axially symmetric atomic nuclei.

\begin{figure}
\includegraphics[width=0.47\textwidth]{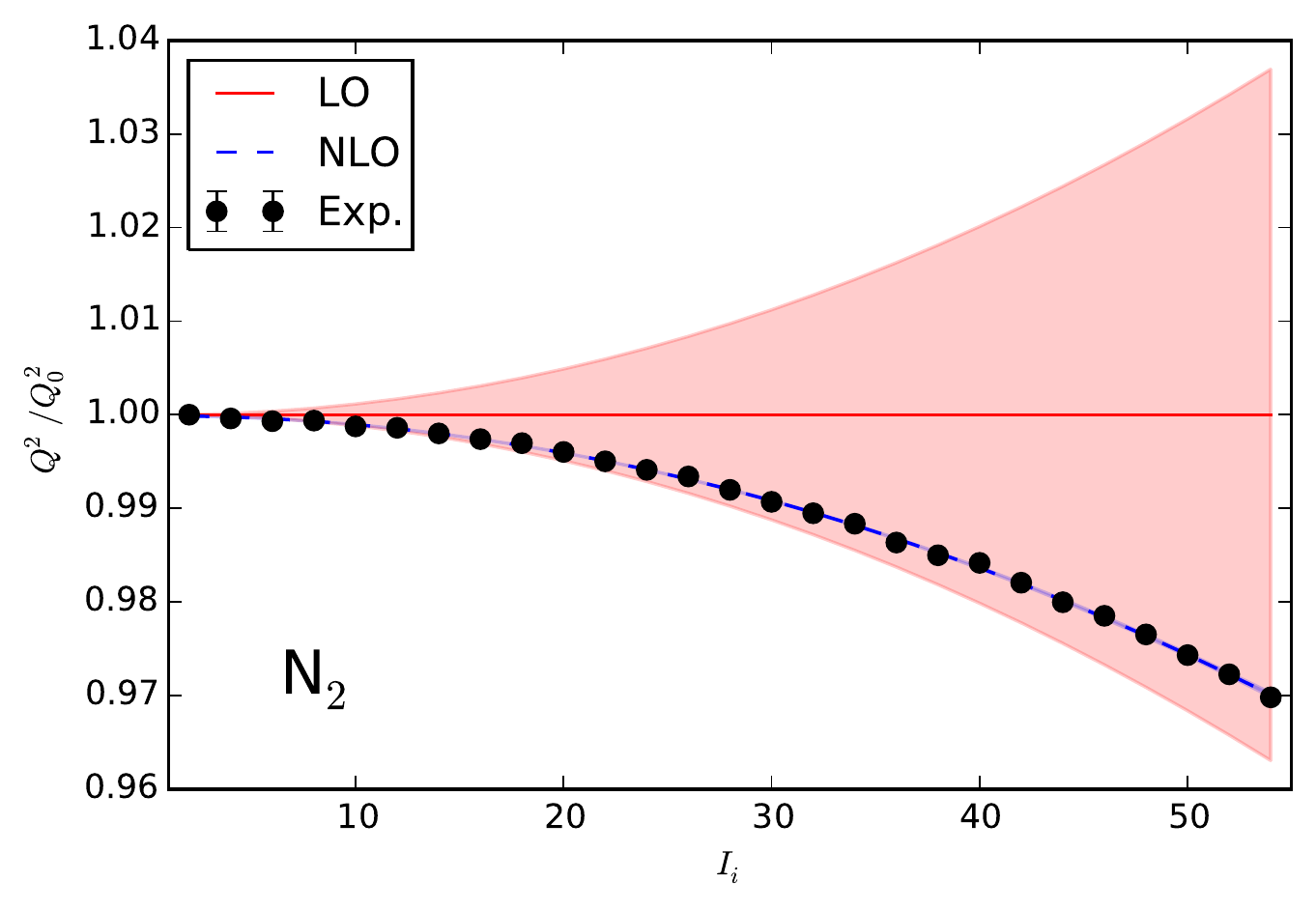}
\caption{Quadrupole transition moments for decays within the ground
  band of the N$_{2}$ molecule in its para state for states with
  initial spin $I_i$. Experimental data~\cite{rothman2013} (black
  circles) is compared to LO (red line and corresponding error band)
  and NLO (blue dashed line and corresponding error band) results of
  the effective theory. The NLO uncertainty band is very small and
  barely visible. The quadratic trend (in spin $I_i$ of the initial
  state) at NLO beyond the constant behavior at LO shows the deviation
  from the rigid rotor.}
\label{be2n2}
\end{figure}

The para N$_{2}$ molecule energy ratios are extremely close to those
of a rigid rotor, see Table~\ref{LECs}. Figure~\ref{be2n2} shows the
experimental data~\cite{rothman2013} of $E2$ transition strengths in
the ground-state band. The LO calculations are in agreement with
experimental data within 1\% for initial angular momenta
$I_{i}\lesssim 30$. NLO calculations deviates from experimental data
less than 0.1\%. The theoretical uncertainty estimates at NLO are
consistent with the data~\footnote{The data~\cite{rothman2013}
  exhibits no experimental uncertainties and we assumed a constant
  error $\sigma_{\rm exp}=0.0002 Q_0$ for a stable fit.}. The values
$\alpha_{\rm LO}$ and $\alpha_{\rm NLO}$ are of natural size, see
Table~\ref{tab:alpha}.

\begin{figure}
\includegraphics[width=0.45\textwidth]{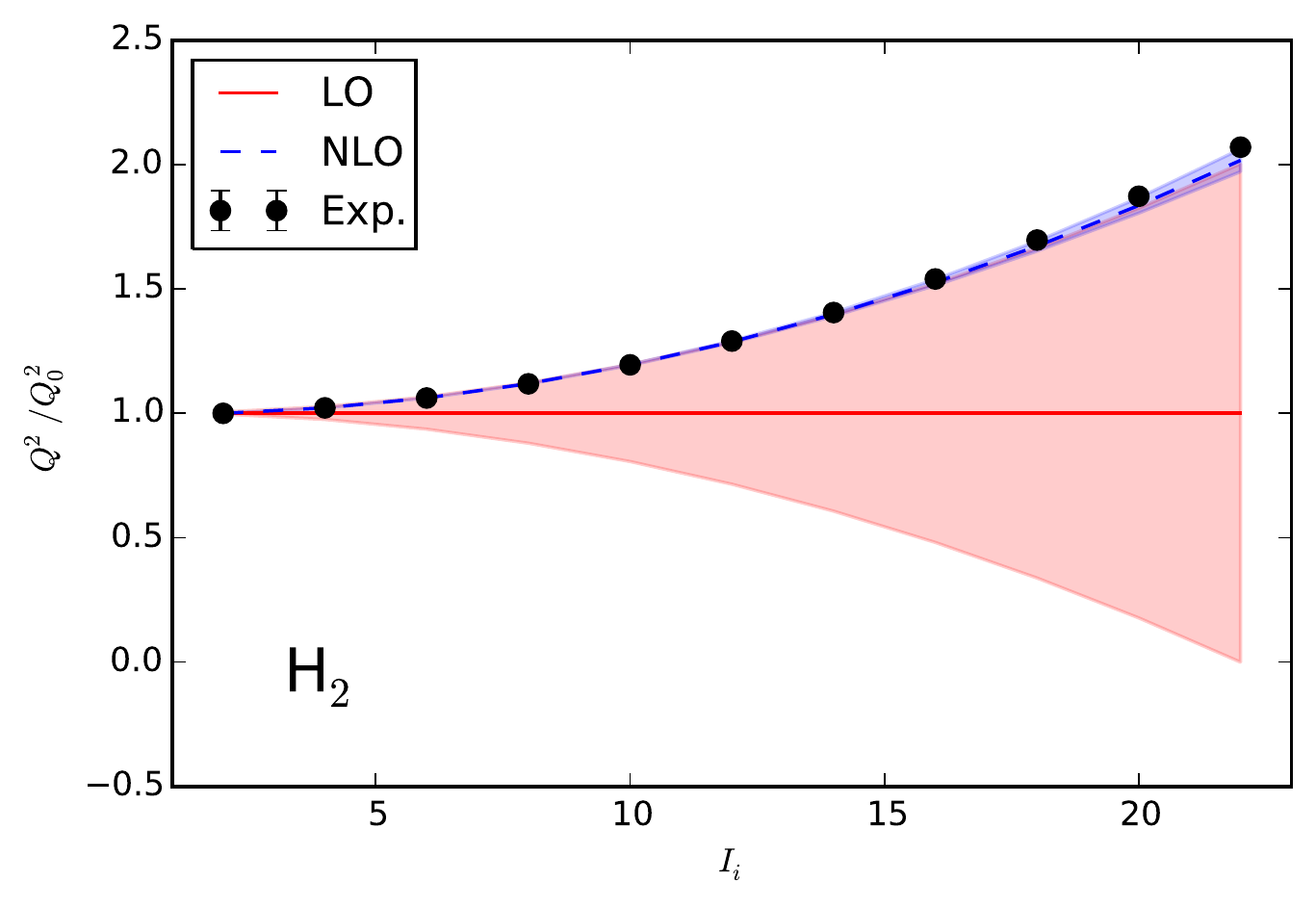}
\caption{Quadrupole transition moments for decays within the ground
  band of the H$_{2}$ molecule in its para state, for states with
  initial spin $I_i$. Experimental data~\cite{rothman2013} (black
  circles) is compared to LO (red line and corresponding error band)
  and NLO (blue dashed line and corresponding error band) results of
  the effective theory.  The quadratic (in spin $I_i$ of the initial
  state) trend at NLO beyond the constant behavior at LO shows the
  deviation from the rigid rotor.}
\label{be2h2}
\end{figure}

The much lighter H$_{2}$ molecule is farther from the rigid-rotor
limit, as shown in Fig.~\ref{be2h2}. Both, LO and NLO calculations are
in agreement with data~\cite{rothman2013}. The value $\alpha_{\rm LO}$
is of natural size, while $\alpha_{\rm NLO}\ll 1$, see
Table~\ref{tab:alpha}. Consequently, the NLO uncertainty is rather
small, possibly because the breakdown scale is at a higher energy than
naively expected. We note that the N$_2$ and H$_2$ molecules
beautifully display that deviations of the quadrupole transitions from
the rigid-rotor limit are quadratic in the spin of the initial state.
This is in accordance with the effective theory.  For the molecules,
the effective theory is accurate (it describes the data) and precise
(theoretical uncertainties are small).

The last column of Table~\ref{LECs} lists the NLO values for the LECs
that enter the quadrupole transition function for the homonuclear
molecules.  Their values are consistent with the NLO correction
$C_2/C_0^3$ obtained from the rotational energy spectrum.

\subsection{Rotational nuclei}
Axially-symmetric deformed nuclei possess positive $\mathcal{R}$
parity, and only states with even angular momentum $I$ are allowed in
the ground-state band.

\begin{figure}
\centering
\includegraphics[width=0.45\textwidth]{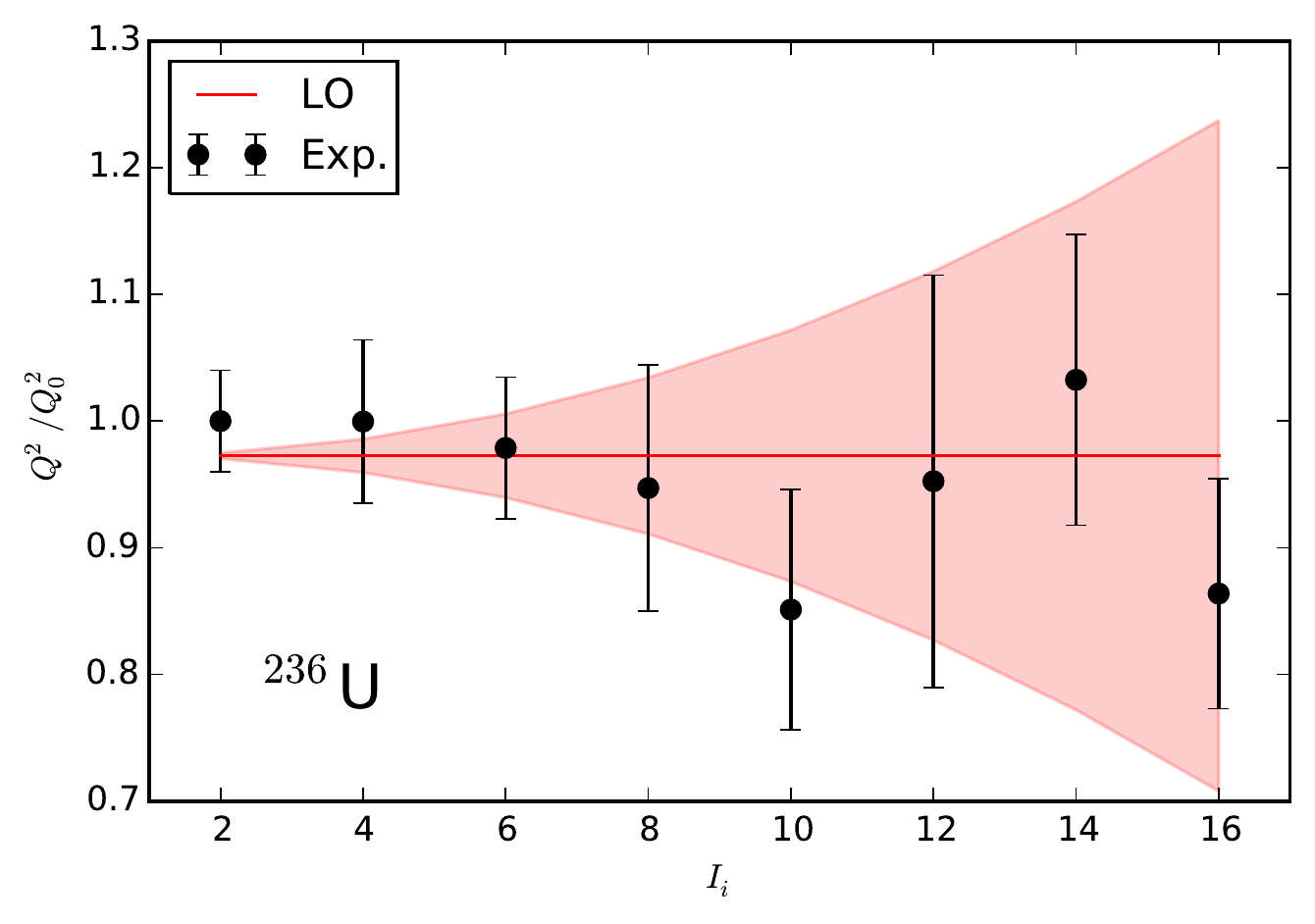}
\caption{Quadrupole decays within the ground-state band of $^{236}$U
  for initial spin $I_i$.  Experimental data~\cite{browne2006}
  with error bars compared to LO calculations of the effective theory.  Estimated
  theoretical uncertainties are shown as bands.}
\label{u236be2}
\end{figure}

The energy spectra of many nuclei in the actinide region makes them
good candidates to test the effective theory. Figure~\ref{u236be2}
shows the quadrupole transition strengths for decays within the ground
band of $^{236}$U and compares them to the experimental data from
\textcite{browne2006}. The results from our LO calculations are in
good agreement with these data. Unfortunately, the experimental
uncertainties are so large that a $\chi^2 <1$ per datum is already
achieved for zero theoretical uncertainties, i.e., for $\alpha_{\rm
  LO}=0$. The shown theoretical uncertainties are obtained
by setting $\alpha_{\rm LO}=1$ for a natural-size estimate.
Data of higher precision would be necessary to probe the theory at
NLO.

\begin{figure}[t!]
\centering
\includegraphics[width=0.45\textwidth]{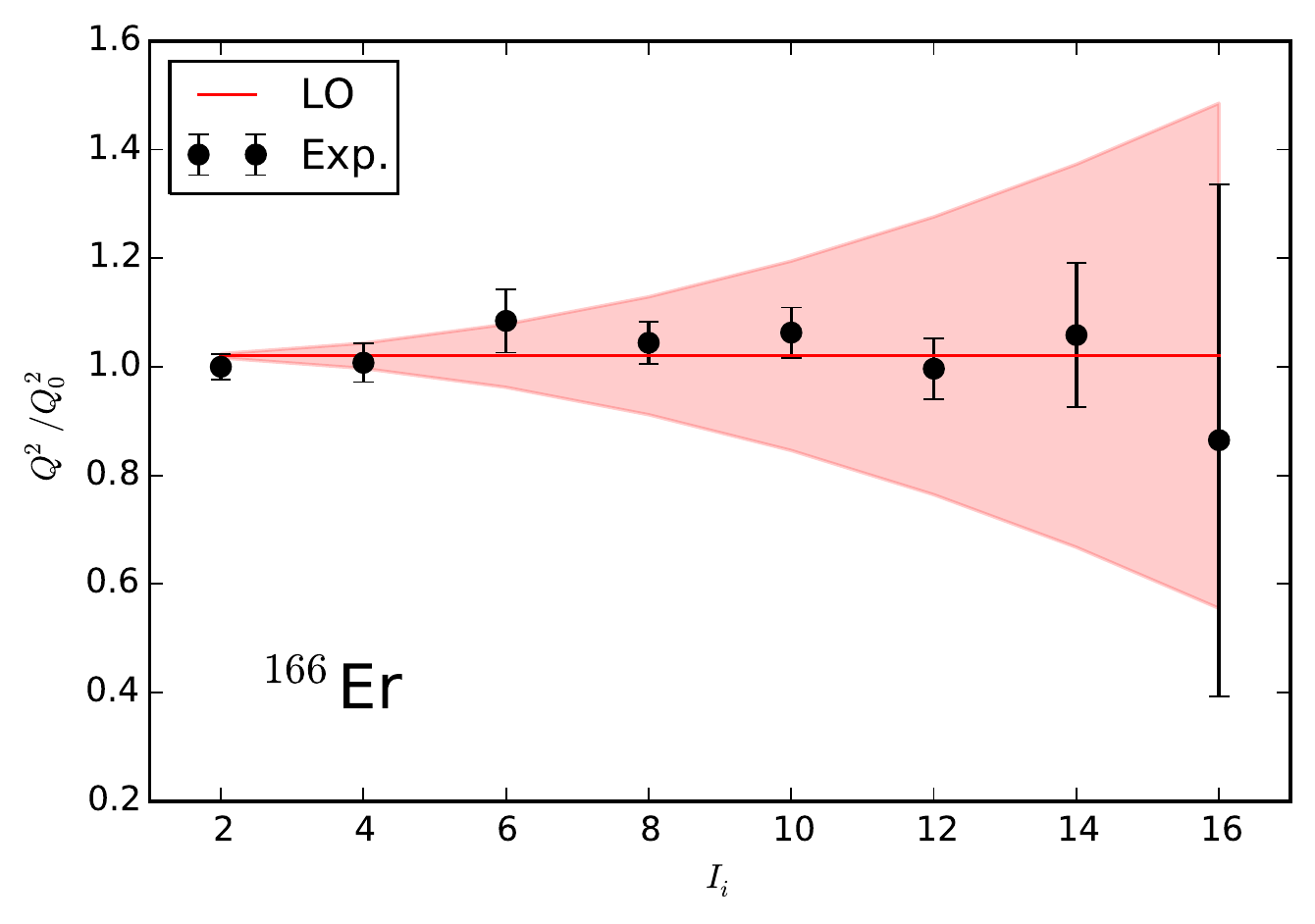}
\includegraphics[width=0.45\textwidth]{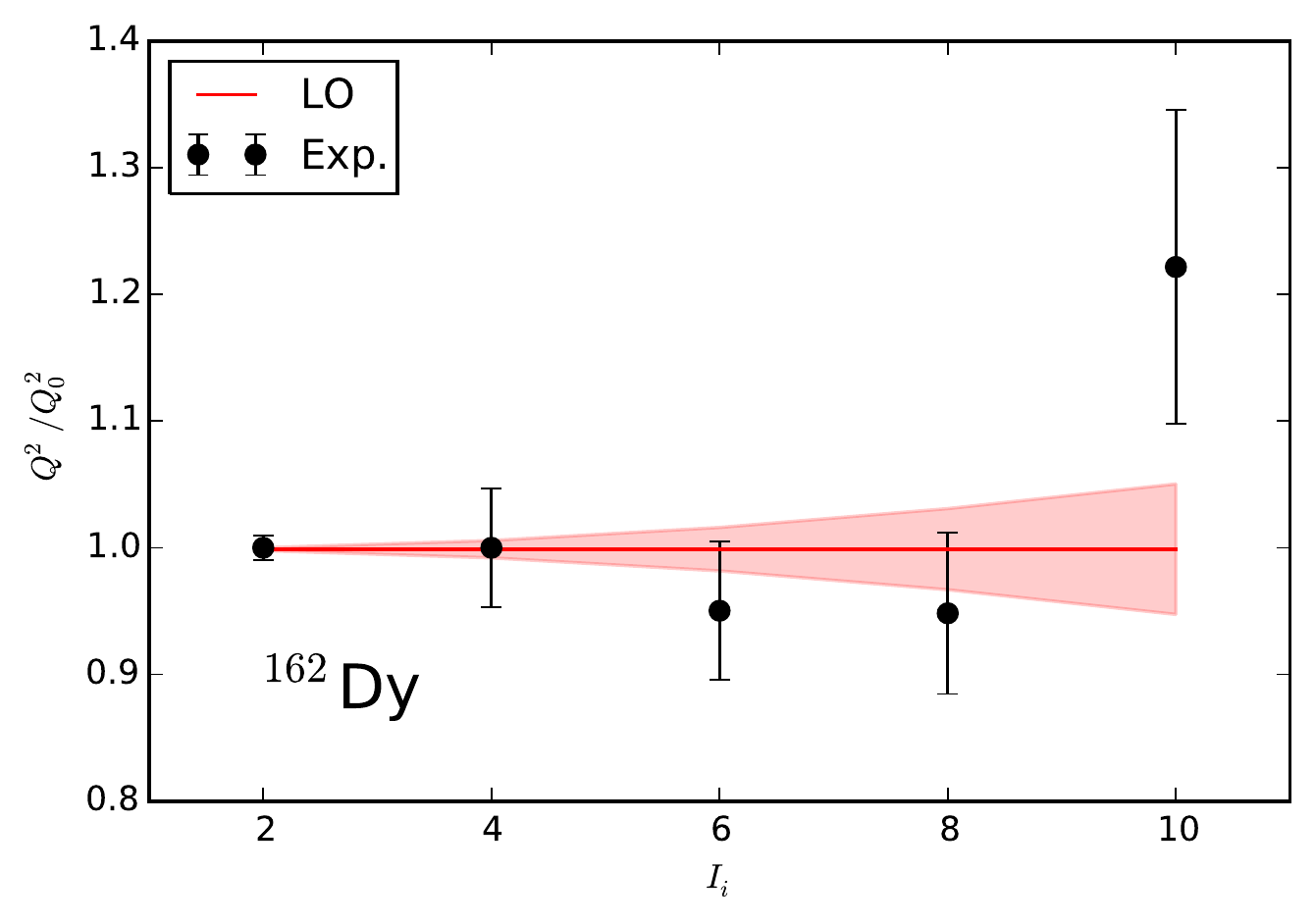}
\caption{Experimental data (black points with error bars) for decays
  within the ground band of $^{166}$Er (top)~\cite{baglin2008} and
  $^{162}$Dy (bottom)~\cite{reich2007} for initial spin $I_i$ is
  compared to LO results (red line with corresponding uncertainty
  band) of the effective theory.  The data is consistent with the
  constant LO value of the effective theory and as expected for a
  rigid rotor.}
\label{rotbe2}
\end{figure}

Many rare-earth nuclei are well deformed, and it is interesting to
confront the effective theory with data. Figure~\ref{rotbe2} shows the
results for the well-studied nuclei
$^{166}$Er~\cite{mcgowan1981,fahlander1992} and
$^{162}$Dy~\citep{hubert1978, aprahamian2006}.  For $^{166}$Er, 
a reduced $\chi^2<1$ is achieved for zero theoretical 
uncertainties (see Table~\ref{tab:alpha}). Same as with
$^{236}$U, the displayed theoretical uncertainties for this 
nucleus employ $\alpha_{\rm LO}=1$ as a natural-size estimate.
For $^{162}$Dy, the data are consistent with the rigid-rotor
result and the error estimates from the effective theory are
natural in size. The first deviation only occurs at higher 
spin, where the experimental uncertainty is increased.
 
\begin{figure}[b!]
\centering
\includegraphics[width=0.45\textwidth]{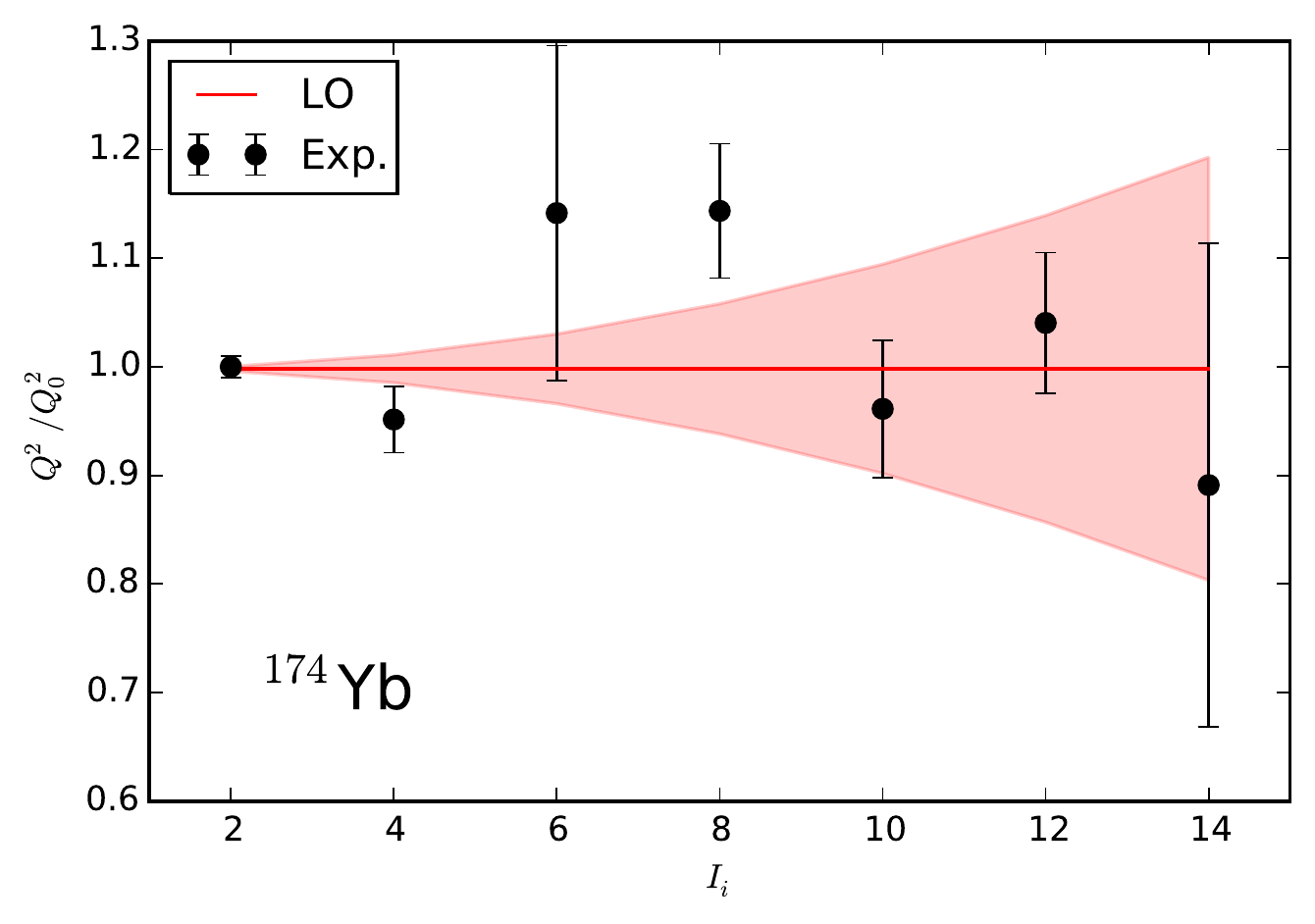}
\includegraphics[width=0.45\textwidth]{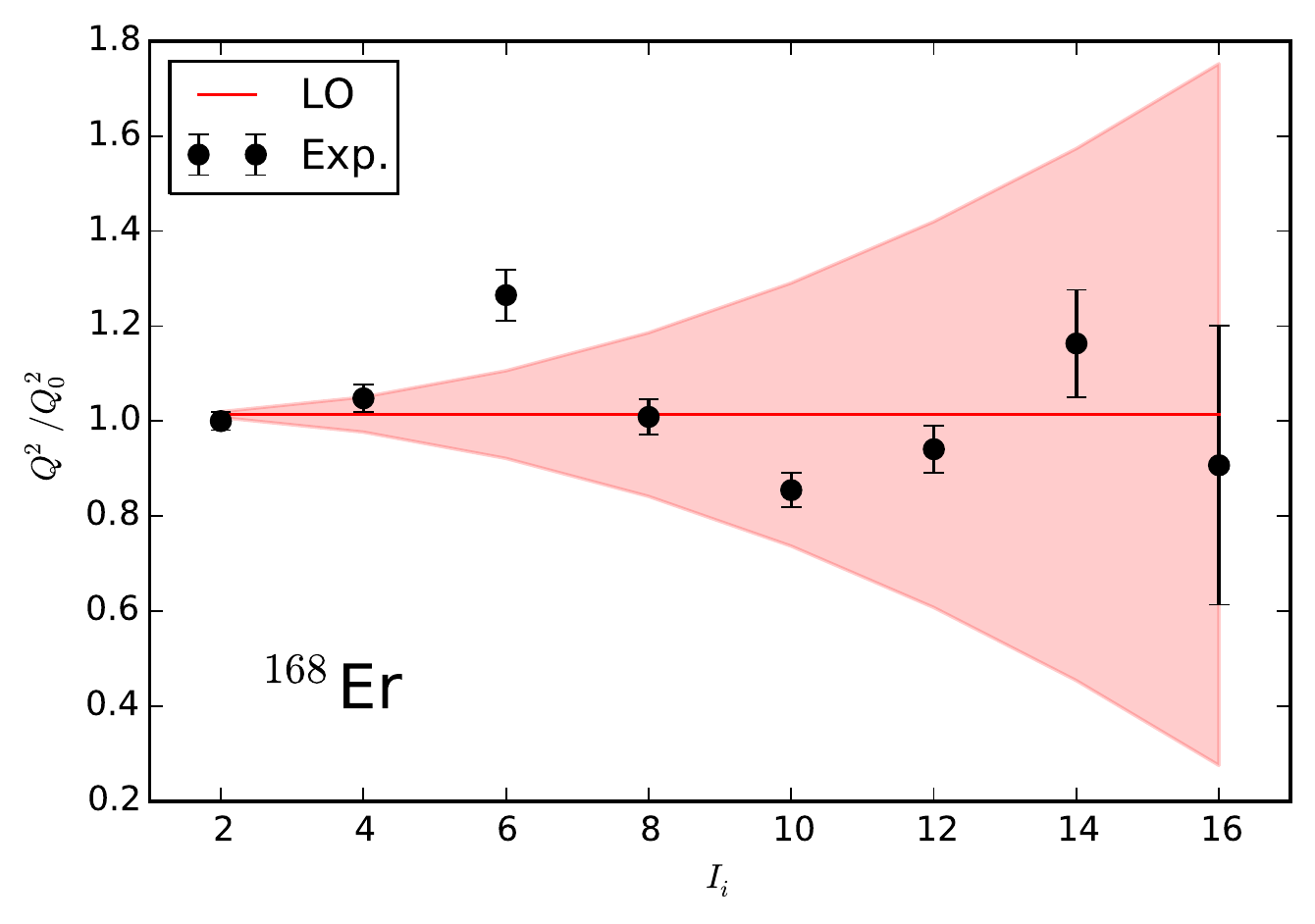}
\includegraphics[width=0.45\textwidth]{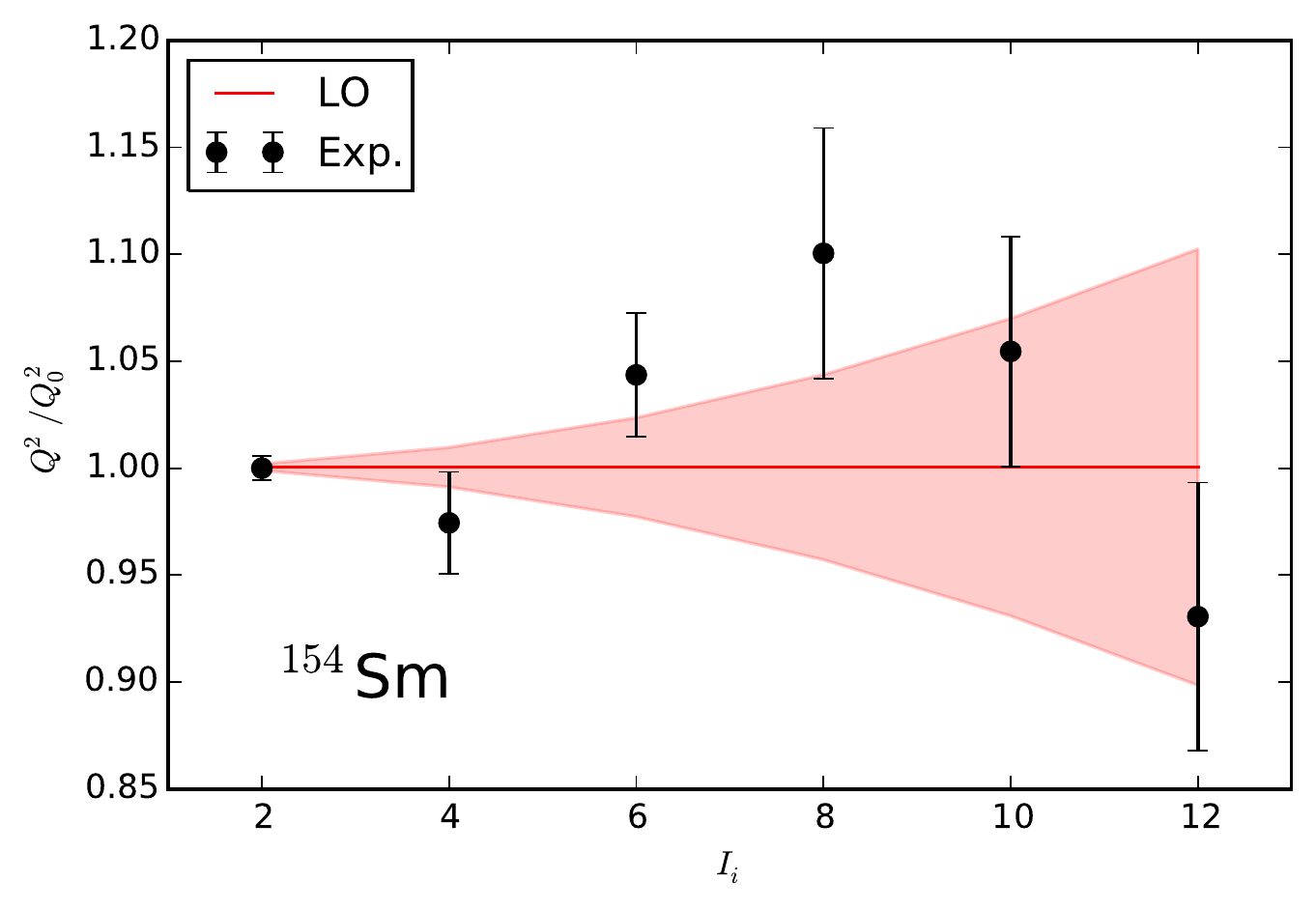}
\caption{Experimental data (black points with error bars) for decays
  within the ground band of $^{174}$Yb (top)~\cite{browne1999},
  $^{168}$Er (middle)~\cite{baglin2010}, $^{154}$Sm
  (bottom)~\cite{reich2009} for initial spin $I_i$ is compared to LO
  results (red line with corresponding uncertainty band) of the
  effective theory. With a few notable exceptions the data is largly
  consistent with the constant LO value of the effective theory and as
  expected for a rigid rotor.  However, the oscillatory pattern of the
  experimental data is not expected within the effective theory.}
\label{rotbe2sm}
\end{figure}

Results for the well deformed nuclei $^{174}$Yb~\cite{browne1999},
$^{168}$Er \cite{davidson1981, warner1981, mcgowan1981,kotlinski1990},
and $^{154}$Sm are shown in Fig.~\ref{rotbe2sm}.  One of the best
rigid-rotor candidates in the rare earth region is $^{174}$Yb due to
its small ratio of $\xi/\omega$. Indeed, the breakdown spin is
conservatively estimated as $\omega/\xi\approx 19$ from the onset of
vibrations and as $\sqrt{C_0^3/C_2}\approx 31$ from the NLO fit to the
spectrum (see Table~\ref{LECs}). The LO results for this nucleus and
our uncertainty estimates are consistent with the experimental
data~\cite{browne1999}.  We note that the data points for the
$4_{g}^{+}\to 2_{g}^{+}$ and the $8_{g}^{+}\to 6_{g}^{+}$ transitions
are below and above the rigid-rotor result $Q=Q_0$. Within the
effective theory, such an oscillatory pattern could only be understood
if the breakdown scale were already around spin $I\approx 6$, and this
is significantly smaller than expected from the ratios $\omega/\xi$ or
$\sqrt{C_0^3/C_2}$ (see Table~\ref{LECs}).  Thus, higher precision
data, particularly for the $6_{g}^{+}\to 4_{g}^{+}$ transition, would
be desirable for this nucleus.

For $^{168}$Er the $6_g^+\to 4_g^+$ transition is significantly away
from the theoretical prediction, and the data exhibit an oscillatory
pattern around the rigid-rotor result. This pattern deviates clearly
from the effective theory's expectation of a deviation quadratic in
initial spin $I_i$ from the rigid-rotor behavior.  Within the
effective theory, such a behavior could only be understood if the
breakdown scale were around the energy of the $6_g^+$ state, which is
unexpectedly low in energy. The relatively large value of $\alpha_{\rm
  LO}$ in Table~\ref{tab:alpha} also reflects the challenge this
nucleus poses. We believe that high-precision measurements,
particularly for the $6_g^+\to 4_g^+$ transition, would be very
interesting for this nucleus.

Finally we turn to $^{154}$Sm. The data is largely consistent with the
rigid rotor results expected at LO in the effective theory. Data
points fall in the very small interval $0.93 Q_0^2\lesssim Q^2\lesssim 1.1
Q_0^2$ around the rigid-rotor prediction.  However, taking the
relatively small experimental error bars at face value would again
suggest that the data oscillates around the constant rigid-rotor
value, and this is not expected within the effective theory.
  
In summary, the data on rotational nuclei is largely consistent with
the LO results that describe a rigid rotor.  A few transition
strengths deviate more than expected from the effective theory, and
one would like to see these data points to be measured with a higher
precision.  In particular, oscillatory patterns around the rigid-rotor
results, as displayed by $^{174}$Yb, $^{168}$Er and possibly
$^{154}$Sm are unexpected and deserve further attention.  The study of
subleading corrections, i.e. deviations expected for a non-rigid
rotor, would require data with considerably higher precision. It is
somewhat surprising that the \citeyear{bohr1975} words of
~\textcite{bohr1975} {\it ``The accuracy of the present measurements
  of $E2$-matrix elements in the ground-state bands of even even
  nuclei is in most cases barely sufficient to detect deviations from
  the leading-order intensity relations''} are still applicable today.
The noted deviations, and the possibility to compare data with more
precise predictions for subleading effects, would make it very
interesting to measure transition strengths in some of these nuclei
with an increased precision.

\subsection{Transitional nuclei}
Transitional nuclei are characterized by energy spectra that deviate
considerably from the rotational behavior. Ratios
$E_{4^+}/E_{2^+}\approx 3$ identify these non-rigid rotors, and the
separation of scale is less pronounced than for the rotational nuclei.
The increased $\xi/\omega$ ratio implies that NLO corrections are more
relevant and also more visible.  Fortunately, for these nuclei data of
sufficiently high precision exists. This allows us to check the
systematic improvements of the effective theory.
 
\begin{figure*}[t!]
\centering
\includegraphics[width=0.45\textwidth]{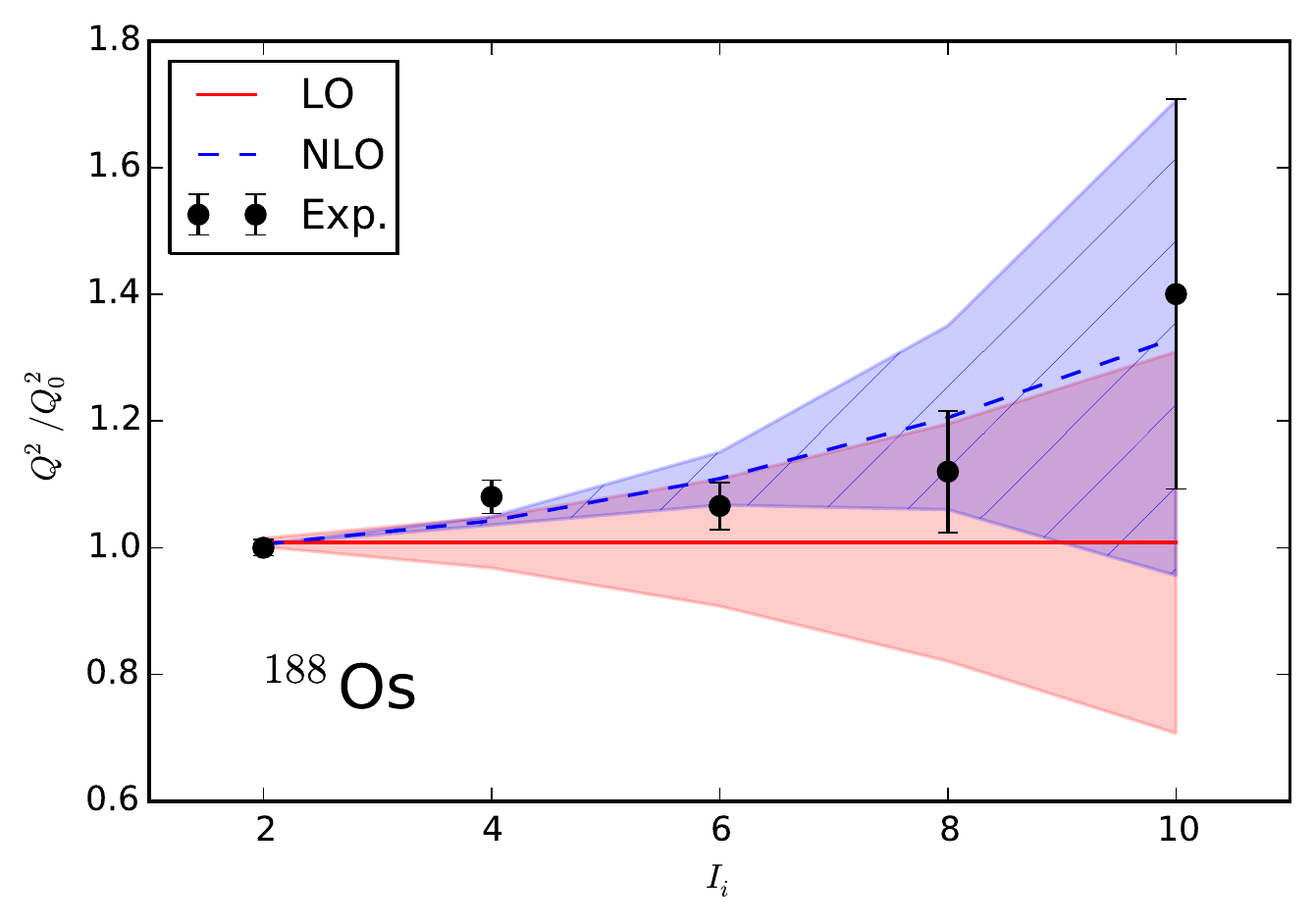}
\includegraphics[width=0.45\textwidth]{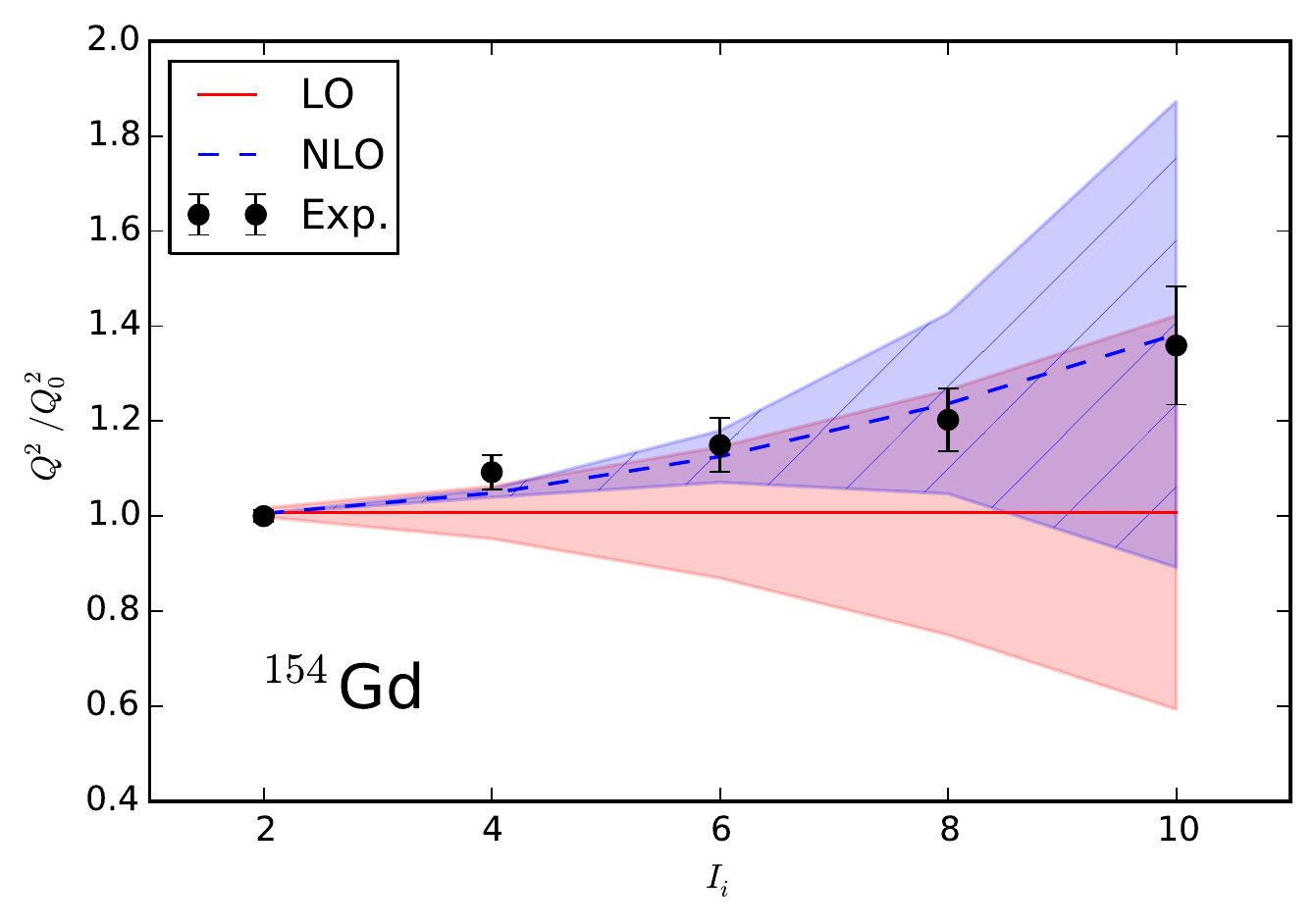}
\includegraphics[width=0.45\textwidth]{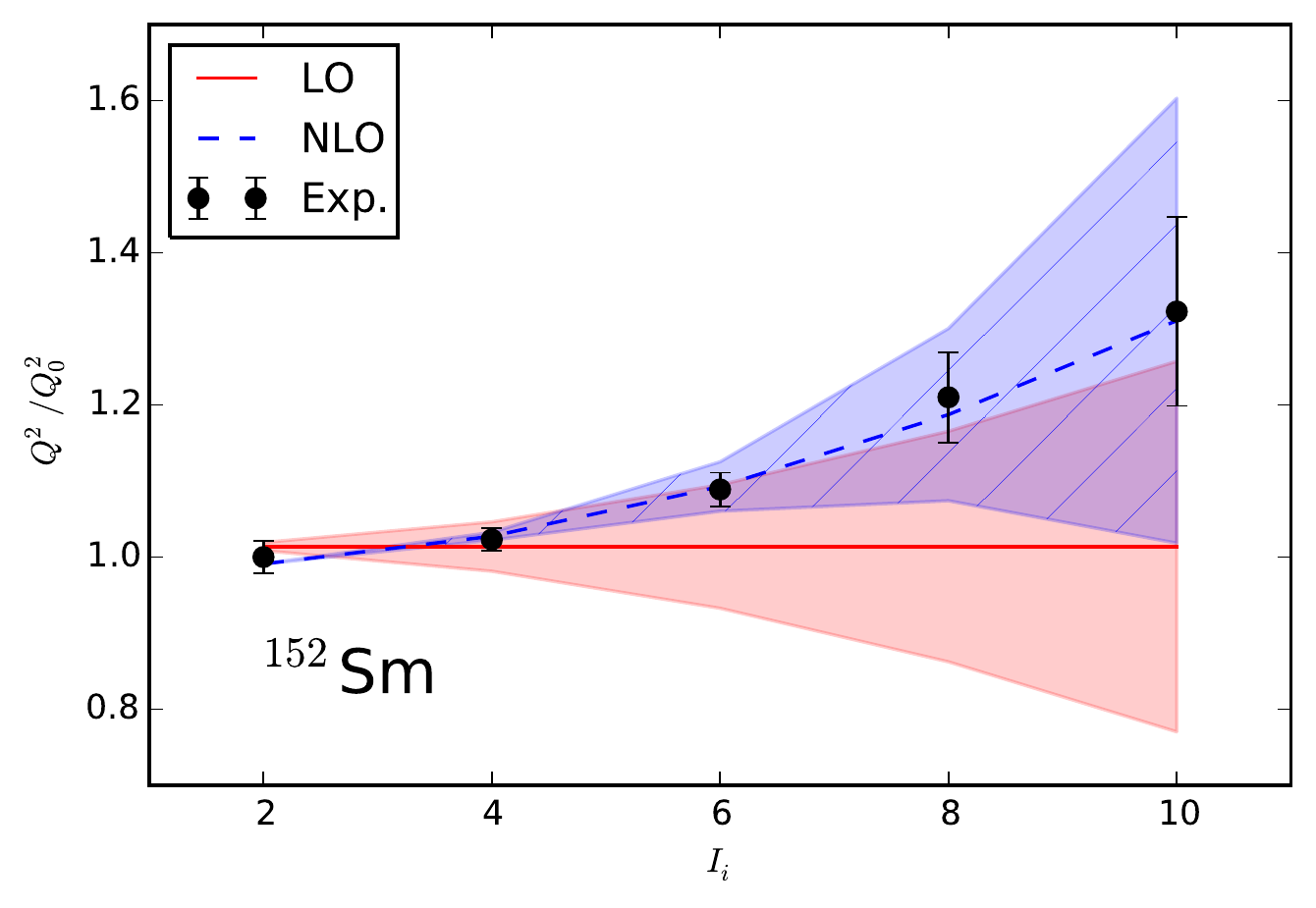}
\includegraphics[width=0.45\textwidth]{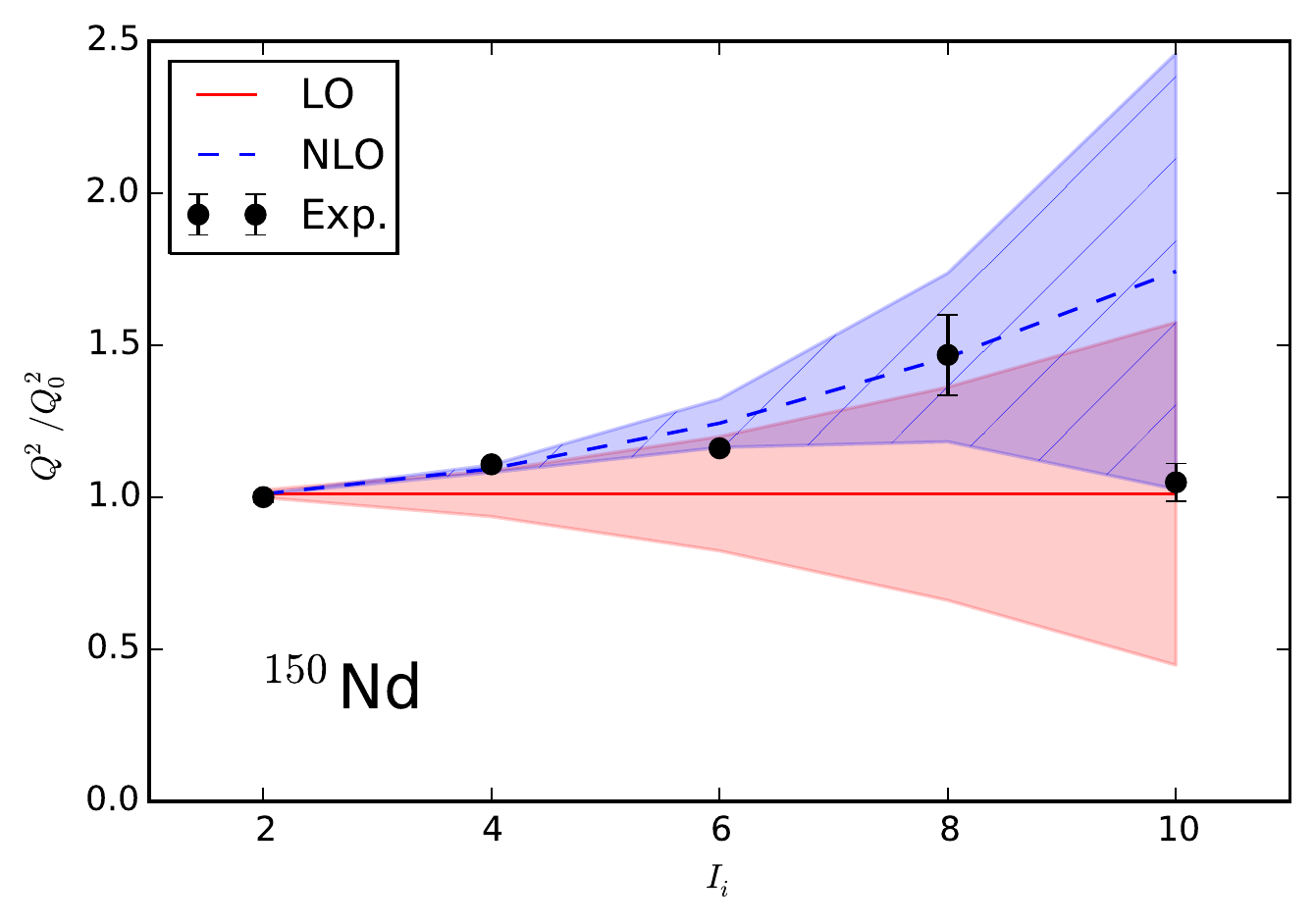}
\caption{Experimental data (black data points with error bars) for
  decays within the ground band of $^{188}$Os (top
  left)~\cite{wu1996b}, $^{154}$Gd (top right)~\cite{tonev2004},
  $^{152}$Sm (bottom left)~\cite{zamfir1999} and $^{150}$Nd (bottom
  right)~\cite{kruecken2002} is compared against LO (red line and
  corresponding uncertainty band) and NLO (blue dashed line with
  corresponding uncertainty band) calculations of the effective
  theory. At NLO, the quadratic deviation (in spin $I_i$) from the LO
  rigid-rotor result is described well by the effective theory.}
\label{be2transitional}
\end{figure*}

Figure~\ref{be2transitional} shows data for quadrupole decays in a few
transitional nuclei and compares them to theoretical results from the
effective theory. For $^{188}$Os (top left panel), the data
systematically deviates from the rigid-rotor result and is
consistently described at LO and at NLO within the theoretical
uncertainties. At spin $I=10$, the theoretical NLO uncertainties
exceed the LO uncertainties, signaling the breakdown of the effective
theory. This is consistent with the expectation
$\sqrt{C_0^3/C_2}\approx 9$ obtained from the fit of the spectrum, see
Table~\ref{LECs}.

The quadrupole transitions of the nucleus $^{154}$Gd (top right panel
of Fig.~\ref{be2transitional}) agree with expectations for a non-rigid
rotor. The quadratic (in $I$) deviations are well described by the
theory at NLO. A $\chi^2<1$ per datum is obtained at NLO even for
vanishing theoretical errors, i.e. for $\alpha_{\rm NLO}=0$ (see
Table~\ref{tab:alpha}).  For the shown NLO error estimates, we set
$\alpha_{\rm NLO}=\alpha_{\rm LO}$.  This choice is of natural size
and consistent with the estimate $\sqrt{C_0^3/C_2}\approx 8$ for the
breakdown spin obtained from the fit of the spectrum, see
Table~\ref{LECs}. The situation is similar for quadrupole transitions
in the ground-state band of $^{152}$Sm (bottom left panel of
Fig.~\ref{be2transitional}). Also here, the shown NLO error estimates
use $\alpha_{\rm NLO}=\alpha_{\rm LO}$.

Finally, we turn to $^{150}$Nd (bottom right panel of
Fig.~\ref{be2transitional}).  This nucleus is a non-rigid rotor and
well described by the LO and NLO effective theory. The relatively
precise value at $I_i=10$ deviates from the quadratic deviation
expected for a non-rigid rotor but is also in the vicinity of the
breakdown scale of the effective theory. Note that the NLO uncertainty
band exceeds the LO uncertainty for $I_i=10$, and this is consistent
with the the estimate $\sqrt{C_0^3/C_2}\approx 7$ for the breakdown
spin obtained from the fit of the spectrum, see Table~\ref{LECs}.

In summary, the effective theory describes the transitional nuclei
rather well. In particular, the quadratic trend (in $I_i$) predicted
as the NLO correction of the effective theory is demonstrated
convincingly.  Theoretical uncertainty estimates are consistent as one
goes from LO to NLO, and they agree with the precision of the
available data. Further progress, e.g the identification of NNLO
corrections, would require even more precise data.  The existing data
suggests that more precise measurements (and possibly the extension to
higher spins) could be particularly profitable for $^{154}$Gd and
$^{152}$Sm.

The successful application of the effective theory to the transitional
nuclei casts further doubts onto the oscillatory patterns in the
experimental data for the rotational $^{174}$Yb and $^{168}$Er, see
Fig.~\ref{rotbe2}. As the breakdown scale for rotational nuclei
considerably exceeds that for transitional nuclei, one would expect
that the effective theory applies even more to the former. This is
additional motivation to re-measure more precisely some of the
critical transitions in well deformed nuclei.

As we have seen, the effective theory allows us to re-derive some of
the well-known results for deformed nuclei~\cite{bohr1975} starting
from symmetry principles alone. New elements are the identification of
a breakdown scale and its employment in a power counting and in
estimates for theoretical uncertainties. In contrast to the
phenomenological models -- which can be accurate -- the effective
theory also delivers precision because it can be improved
systematically.  It is also encouraging that well deformed and
transitional nuclei are described on the same footing, without
resorting to more special models~\cite{iachello2001} for the latter.
For the results presented in this Section, the predictive power of the
effective theory equals the traditional approaches~\cite{bohr1975}. At
LO, one LEC is used to describe the spectrum, and one describes the
quadrupole transition strengths. At NLO, one additional LEC each
enters the spectrum and the transitions.

\section{Rotations and vibrations}
\label{Vibrations}
In this Section, we are interested in inter-band transitions. These
transitions are much weaker than the intra-band transitions considered
in the previous Section, and the accurate description of these faint
transitions poses a challenge.  For the description of rotational
bands beyond the ground-state band, we need to include additional
degrees of freedom into the effective theory. For even-even nuclei,
these degrees of freedom represent higher-energetic vibrations of the
nucleus with an energy scale $\omega$ below the breakdown energy scale $\Lambda$.
These vibrations are the true remnants of Nambu-Goldstone modes in
finite systems with emergent symmetry breaking~\cite{papenbrock2014}.
The effective theory for this case has in parts been developed in
Refs.~\cite{papenbrock2011,zhang2013}.
Reference~\cite{papenbrock2011} developed the effective theory up to
NLO. In leading order, the theory describes uncoupled vibrational
states. At NLO, the vibrational states become heads of rotational
bands. Reference~\cite{zhang2013} focused on higher-order terms. Then,
couplings between vibrational band heads enter, and the dependence of
the moment of inertia on the quantum numbers of the band heads could
be described.  In the next Subsection, we briefly
introduce quadrupole degrees of freedom. We then develop the
Hamiltonian up to NNLO, and finally focus on the coupling of electromagnetic
fields for the description of inter-band transitions in the following Section.

\subsection{Quadrupole degrees of freedom}
In even-even nuclei, the quantized Nambu-Goldstone modes due to the
emergent symmetry breaking from SO(3) to SO(2) can be represented by a
quadrupole field with two of its components replaced by the low-energy
degrees of freedom $\theta$ and $\phi$.  Note that these degrees of
freedom have the quantum numbers of quadrupole modes, but they are not
Bohr's surface oscillations. In our treatment, the quadrupole degrees
of freedom also realize the rotational symmetry nonlinearly; they are
in the co-rotating coordinate system and can be viewed as being
attached to the particle moving on the two-sphere. We have
\begin{equation}
\Psi=\left(\Psi_{+2}, 0, \Psi_{0}, 0, \Psi_{-2}\right).
\end{equation}
It is convenient to rewrite the components of the field as
\begin{equation}
\Psi_{0}=\zeta+\psi_{0}\qquad\Psi_{\pm 2}=\psi_{2}e^{\pm i2\gamma} \ .
\end{equation}
Here $\zeta$ is the constant vacuum expectation value of the zero mode
(with $|\psi_0| \ll \zeta$), and the factor 2 in the phase $2\gamma$
has been introduced for convenience. Because of this factor $\gamma$
ranges from $0$ to $\pi$.

The field in the laboratory frame can be written as an appropriate
rotation of the field in the intrinsic frame
\begin{equation}
\Phi=g(\phi,\theta)\Psi,
\end{equation}
which implies that under the SO(3) rotation $r(\alpha,\beta,\gamma)$
\begin{equation}
\psi_0\rightarrow\psi_0\qquad\psi_2\rightarrow\psi_2\qquad\gamma\rightarrow\gamma+\chi.
\end{equation}
Here, $\chi=\chi(\alpha,\beta,\gamma;\theta,\phi)$
is a complicated function of the rotation angles and the orientation
angles $(\theta,\phi)$; the rotational symmetry is realized
nonlinearly~\cite{papenbrock2011}.

The kinetic terms in the quadrupole degrees of freedom are obtained by
acting with the covariant derivative (\ref{Dt}) onto $\Psi$. Thus, any
Lagrangian $L$ in $v_{\pm}$, $\Psi_{0}$, $\Psi_{\pm 2}$,
$D_{t}\Psi_{0}$, $D_{t}\Psi_{\pm2}$ that is formally invariant under
SO(2), is invariant under SO(3) due to the nonlinear realization of
the rotational symmetry. The application of Noether's theorem to such
Lagrangian yields the total angular momentum $\mathbf{J}$ with
spherical components
\begin{equation}
\label{Itot}
\begin{split}
J_{+1} &= -{1\over\sqrt{2}}e^{i\phi}(ip_{\theta}-p_{\phi}\cot{\theta})-{\frac{1}{\sqrt{2}}}e^{i\phi}\frac{p_{\gamma}}{\sin{\theta}} \\
J_{0} &= p_{\phi} \\
J_{-1} &= -{1\over\sqrt{2}}e^{-i\phi}(ip_{\theta}+p_{\phi}\cot{\theta})+\frac{1}{\sqrt{2}}e^{-i\phi}\frac{p_{\gamma}}{\sin{\theta}},
\end{split}
\end{equation}
as the conserved quantity. Here
\begin{equation}
p_{\theta}\equiv\partial_{\dot{\theta}}L\qquad p_{\phi}\equiv\partial_{\dot{\phi}}L\qquad p_{\gamma}\equiv\partial_{\dot{\gamma}}L,
\end{equation}
and the total angular momentum squared is
\begin{equation}
\mathbf{J}^{2}=p_{\theta}^{2}+\left(\frac{p_{\phi}-p_{\gamma}\cos{\theta}}{\sin{\theta}}\right)^{2}+p_{\gamma}^{2}.
\end{equation}
We denote the total angular momentum as $\mathbf{J}$, because its definition
(\ref{Itot}) differs from Eq.~(\ref{spin}) due to the newly introduced
vibrational degrees of freedom.

Let us briefly compare the degrees of freedom in the effective theory
to those of the Bohr Hamiltonian. In the effective theory, the angles
$(\phi,\theta,\gamma)$ can be viewed as Euler angles, with the
``slow'' degrees of freedom $\Omega=(\theta,\phi)$ describing the
orientation of the symmetry axis and the ``fast'' degree of freedom
$\gamma$ describing rotations around the symmetry axis. The $\psi_0$
degree of freedom is a (fast) vibration that keeps the axial symmetry,
while $\psi_2$ is a (fast) vibration that breaks the axial symmetry.
The Bohr Hamiltonian employs three Euler angles and two deformation
parameters~\cite{rohozinski2009}. The two deformation parameters
(usually labeled as $\beta$ and $\gamma$, respectively) describe the
amplitude of the total deformations ($\beta$), and the deformation
breaks axial symmetry for $\gamma\ne 0$. The variable $\beta$ can be
viewed as a hyper radius in five-dimensional space of the quadrupole
degrees of freedom, while $\gamma$ is a hyper angle in addition to the
three Euler angles.
 
\subsection{Power counting and Hamiltonian at NNLO}
In addition to the power counting estimates~(\ref{power}) we
have~\cite{papenbrock2011}
\begin{equation}
\begin{split}
\omega_0\sim\omega_2\sim\dot{\gamma}\sim\omega &\qquad \dot{\psi_{0}}\sim\dot{\psi_{2}}\sim\omega^{1/2}\\
\zeta\sim\xi^{-1/2} &\qquad \psi_{0}\sim\psi_{2}\sim\omega^{-1/2}.
\end{split}
\label{fullpower}
\end{equation}
For an understanding of this scaling we recall that the angles
$\theta$, $\phi$ and $\gamma$ are dimensionless, and that a time
derivative on these degrees of freedom must scale as the excitation
energy of the motion they generate. The scaling of $\psi_{i}$,
$i=0,2$, is such that $\dot{\psi_{i}}^{2}\sim\omega$. The expectation
value $\zeta$ is associated with the emergent symmetry breaking and
must scale as $\xi^{-1/2}$.

The Lagrangian of this effective theory is $L_{\rm LO}+L_{\rm
  NLO}+L_{\rm NNLO}$ where the leading-order Lagrangian
\begin{equation}
L_{\rm LO}=\frac{1}{2}\dot{\psi}_{0}^{2}+\dot{\psi}_{2}^{2}+4\dot{\gamma}^{2}\psi_{2}^{2}-\frac{\omega_{0}^{2}}{2}\psi_{0}^2-\frac{\omega_{2}^{2}}{4}\psi_{2}^2
\end{equation}
describes vibrations at the high-energy scale $\omega$, the NLO
correction
\begin{equation}
L_{\rm NLO}=\frac{C_{0}}{2}\left(\dot{\theta}^2+\dot{\phi}^2\sin^2{\theta}\right)+4\psi_{2}^{2}\dot{\gamma}\dot{\phi}\cos{\theta}
\end{equation}
couples rotations at the low-energy scale $\xi$ to vibrations via the
$\gamma$ degree of freedom, and the NNLO correction
\begin{equation}
\begin{split}
L_{\rm NNLO} &= \frac{C_{\beta}}{2}\psi_{0}\left(\dot{\theta}^2+\dot{\phi}^2\sin^2{\theta}\right)\\
&+ \frac{C_{\gamma}}{2}\psi_{2}\left(\dot{\theta}^2-\dot{\phi}^2\sin^2{\theta}\right)\cos{2\gamma}\\
&+ C_{\gamma}\psi_{2}\dot{\theta}\dot{\phi}\sin{2\gamma}\sin{\theta},
\end{split}
\end{equation}
is treated as a perturbation that scales as $\xi(\xi/\omega)^{1/2}$.
According to the power counting~\ref{fullpower}, this implies
\begin{equation}
C_{\beta}\sim C_{\gamma}\sim\xi^{-1/2}.
\end{equation}
Note that $\gamma$ is a cyclic variable of the LO and NLO Lagrangians.
Thus, at these orders, the projection of the angular momentum
$\mathbf{J}$ onto the intrinsic symmetry axis $p_{\gamma}$, is a
conserved quantity in addition to the total angular
momentum~(\ref{Itot}).

A Legendre transformation of the Lagrangian yields the Hamiltonian
$H_{\rm LO}+H_{\rm NLO}+H_{\rm NNLO}$. Here
\begin{equation}
H_{\rm LO}=\frac{p_{0}^{2}}{2}+\frac{\omega_{0}^{2}}{2}\psi_{0}^{2}+\frac{p_{2}^{2}}{4}+\frac{1}{4\psi_{2}^{2}}\left(\frac{p_{\gamma}}{2}\right)^{2}+\frac{\omega_{2}^{2}}{4}\psi_{2}^{2}
\end{equation}
is the Hamiltonian of a harmonic oscillator with frequency
$\omega_{0}$ coupled to a two-dimensional harmonic oscillator with
frequency $\omega_{2}$. The quantization is standard
\begin{equation}
\hat{p}_{0}=-i\partial_{\psi_{0}}\qquad\hat{p}_{2}=-i\partial_{\psi_{2}}\qquad\hat{p}_{\gamma}=-i\partial_{\gamma}.
\end{equation}
We denote the eigenstates of the LO Hamiltonian as
$|n_{0}n_{2}K/2\rangle$, with integer $n_0$ and $n_2$ and even $K$.
Here $n_{0}$, $n_{2}$ and $K/2$ are the number of quanta of the modes
$\psi_0$, $\psi_2$, and $\gamma$, respectively. These states can be
written as $|n_{0}\rangle|n_{2}\rangle|K/2\rangle$, where
$|n_{0}\rangle$ are the states of the harmonic oscillator, and
$\langle\psi_{2}|n_{2}\rangle$ are the radial wave functions of the
two-dimensional harmonic oscillator.

The NLO correction
\begin{equation}
H_{\rm NLO}=\frac{1}{2C_{0}}\mathbf{p}_{\Omega\gamma}^{2}=\frac{1}{2C_{0}}\left(\mathbf{J}^2-p_\gamma^2\right)
\end{equation}
is the Hamiltonian of a symmetric top~\cite{merzbacher1961}. Here, the
momentum in the tangential plane is
\begin{equation}
\mathbf{p}_{\Omega\gamma}=\mathbf{e}_{\theta}p_\theta+\mathbf{e}_{\phi} p_{\phi\gamma}, 
\end{equation}
with
\begin{equation}
p_{\phi\gamma}\equiv\frac{p_{\phi}-p_{\gamma}\cos{\theta}}{\sin{\theta}}.
\end{equation}
We also have
\be
\mathbf{J}=\mathbf{e}_r\times\mathbf{p}_{\Omega\gamma} + \mathbf{e}_r p_\gamma \ .
\ee
This form of the angular momentum agrees with the intuition. In
particular, rotations around the symmetry axis $\mathbf{e}_r$ yield a
contribution to the angular momentum in the direction of this axis.

The quantization
\begin{equation}
\begin{split}
\hat{\mathbf{p}}_{\Omega\gamma} &= -i\mathbf{e}_{\theta}\partial_{\theta}-i\mathbf{e}_{\phi}\frac{\partial_{\phi}-\partial_{\gamma}\cos{\theta}}{\sin{\theta}}\\
&=-\mathbf{e}_{r}\times\hat{\mathbf{J}}
\end{split}
\end{equation}
is standard.  In what follows, we denote the differential operator
corresponding to the momentum operator $\hat{\mathbf{p}}_{\Omega\gamma}$
also as
\be
-i\mathbf{\nabla}_{\Omega\gamma} \equiv \mathbf{p}_{\Omega\gamma}. 
\ee
The Hamiltonian eigenvalue problem becomes 
\begin{equation}
\hat{H}_{\rm NLO}|IMK\rangle=\frac{1}{2C_{0}}\left[I(I+1)-K^2\right]|IMK\rangle.
\end{equation}
Here, we continued to denote the eigenvalues of the total angular
momentum by the quantum number $I$. The wave functions are linear
combinations of Wigner $D$ functions, consistent with the positive
$\mathcal{R}$ parity of the system, i.e.
\begin{equation}
\langle\Omega\gamma|IMK\rangle=N\left[D_{MK}^{I}(\Omega,\gamma)+(-1)^{I}D_{M-K}^{I}(\Omega,\gamma)\right].
\end{equation}
Here $N$ is a normalization factor. For $K=0$, the wave function
cannot take odd $I$ values due to the $\mathcal{R}$ parity. Thus, for
even $I$ the wave function takes the form
\begin{equation}
\langle\Omega\gamma|IM0\rangle=\sqrt{\frac{2I+1}{4\pi^{2}}}D_{M0}^{I}(\Omega,\gamma)=\frac{(-1)^{m}}{\sqrt{\pi}}Y_{I-M}(\Omega).
\end{equation} 
The Wigner D-functions $D_{MK}^{I}(\Omega\gamma)$ fulfill the
relations~\cite{varshalovich1988}
\begin{equation}
\begin{split}
\hat{J}_{z}D_{MK}^{J}(\Omega\gamma) &= -MD_{MK}^{I}(\Omega\gamma)\\
\hat{J}_{z'}D_{MK}^{J}(\Omega\gamma) &= -KD_{MK}^{I}(\Omega\gamma)\\
\hat{\mathbf{J}}^{2}D_{MK}^{J}(\Omega\gamma) &= I(I+1)D_{MK}^{I}(\Omega\gamma).
\end{split}
\end{equation}

The complete Hamiltonian at NLO can be diagonalized as
\begin{eqnarray}
\label{eigen_nlo}
\lefteqn{\left(\hat{H}_{\rm LO}+ \hat{H}_{\rm NLO}\right)|n_{0}n_{2}IMK\rangle =
\bigg[ \omega_{0}\left(n_{0}+\frac{1}{2}\right) }\\
&&+ {\omega_{2}\over 2}\left(2n_{2}+\frac{K}{2}+1\right)+ \frac{I(I+1)-K^2}{2C_0}\bigg] |n_{0}n_{2}IMK\rangle.\nonumber
\end{eqnarray}
Thus, at this order, the spectrum consists of rotational bands with
rotational constant $1/(2C_{0})$ on top of harmonic vibrations. The
vibrational quanta determine the band head, and the ground-state band
has no vibrational quanta excited. Because $0\le \gamma \le \pi$, the wave
function in $\gamma$ must exhibit periodic boundary conditions at the
domain boundaries. This limits $K$ to even values.  Historically, the
band head with $(n_{0}=1, n_2=0, K=0)$, and the band head with
$(n_0=0,n_2=0,K=2)$ determine the ``$\beta$ band'' and the ``$\gamma$ band''
respectively. In what follows, we continue to use these labels. 

The NNLO correction to the Hamiltonian is
\begin{equation}
\label{hamNNLO}
H_{\rm NNLO}=-\frac{1}{2C_{0}^{2}}\left(C_{\beta}\psi_{0}\mathbf{p}_{\Omega\gamma}^{2}+C_{\gamma}\psi_{2}\mathbf{p}_{\Omega\gamma}^{T}\hat{\Gamma}\mathbf{p}_{\Omega\gamma}\right).
\end{equation}
Here,
\begin{equation}
\label{Bmat}
\hat{\Gamma}\equiv \left[\begin{array}{c c}
\cos{2\gamma} & \sin{2\gamma}\\
\sin{2\gamma} & -\cos{2\gamma}\\
\end{array}\right] 
\end{equation}
acts on vectors in the tangent plane. The operator $\hat{H}_{\rm
  NNLO}$ is off-diagonal in the eigenstates of the NLO Hamiltonian.
Thus, it is only effective in second-order perturbation theory, i.e.
at order N$^3$LO. At that order, corrections to the rotational
constant (or the effective moment of inertia) linear in the number of
excited quanta are introduced~\cite{zhang2013}. These corrections
arise due to omitted physics at the breakdown scale $\Lambda\sim
3$~MeV, where pair-breaking effects need to be taken into
account~\cite{papenbrock2014}. Thus, deviations from the harmonic
behavior of the band heads is expected to scale as
$\omega/\Lambda\approx 1/3$ for nuclei in the rare-earth and actinide
regions. In the following Section, we will determine the LECs
$C_\beta$ and $C_\gamma$ by fit to inter-band transitions. In the long
run, it would be interesting to compute LECs from more microscopic
methods~\cite{caprio2013,dytrych2013}.

\section{Inter-band transitions}
\label{Interband}
In this Section, we couple electromagnetic fields to the Hamiltonian,
and focus on the inter-band transitions. These transitions are much
fainter than the strong intra-band transitions discussed in the
Sect.~\ref{Intraband}. The transitions from the $\beta$ band to the
ground-state band are not understood very well (see
Ref.~\cite{garrett2001} for a review), because the traditional
models overpredict them by up to an order of magnitude. Furthermore,
these transitions vary by about two orders of magnitude in
well-deformed and transitional nuclei~\cite{aprahamian2004}.  Below we
will see that the transitions pose no challenge to the effective
theory.  For the LO description of these transitions, we only need to
gauge the NNLO Hamiltonian.

\subsection{Transition operators}
The NNLO Hamiltonian of the previous section can be coupled to an
electromagnetic field employing the gauging
\begin{equation}\label{fullgauging}
\hat{\mathbf{p}}_{\Omega\gamma}\rightarrow\hat{\mathbf{p}}_{\Omega\gamma}-q\mathbf{A}_\Omega=-i\nabla_{\Omega\gamma}-q\mathbf{A}_\Omega \ . 
\end{equation}
This is equivalent to  
\be
\mathbf{J}\to\mathbf{J}-q\mathbf{e}_r\times\mathbf{A}_\Omega \ , 
\ee
and in full analogy to Eq.~(\ref{gauging}). 

Thus, the angles $\theta$, $\phi$, and $\gamma$ are gauged. Assuming
that the vibrational degrees of freedom $\psi_0$ and $\psi_2$ also
carry a charge, we could also couple these to the radial component
$\mathbf{A}\cdot\mathbf{e}_r$ to obtain a rotationally invariant and
gauge-invariant Hamiltonian. As discussed below, the corresponding
terms do not yield independent contributions for the intra-band
transitions considered in this paper, and they are therefore
neglected.

The gauging of the NNLO contribution~(\ref{hamNNLO}) to the Hamiltonian 
\bea
\label{ham_nnlo_A}
\hat{H}_{\rm NNLO}^{(\mathbf{A})} &=& \frac{iq}{2C_{0}}\frac{C_{\beta}}{C_{0}}\psi_{0}\left(\mathbf{A}\cdot\nabla_{\Omega\gamma} + \nabla_{\Omega\gamma}\cdot\mathbf{A}\right)\nonumber\\
&+& \frac{iq}{2C_{0}}\frac{C_{\gamma}}{C_{0}}\psi_{2}\left(\mathbf{A}^{T}\hat{\Gamma}\nabla_{\Omega\gamma} + \nabla_{\Omega\gamma}^{T}\hat{\Gamma}\mathbf{A}\right)
\eea
induces LO inter-band transitions. As the inter-band transitions
originate from a small correction to the Hamiltonian, they are
expected to be an order of magnitude weaker (in the power counting)
than the intra-band transitions. Gauging of the fields $\psi_0$ and
$\psi_2$ would add terms $q_0A_r \hat{p}_0$ and $q_2 A_r\hat{p}_2$ to
the Hamiltonian. Here $A_r=\mathbf{A}\cdot\mathbf{e}_r$. These
operators do not yield transition matrix elements that differ from
those of the operators in the Hamiltonian~(\ref{ham_nnlo_A}). They are
therefor neglected.

Following Eq.~(\ref{be2}) we compute the transition strength as
\be
B(E\lambda,i\rightarrow f)=\frac{1}{2l_{i}+1}\left|\langle f||\mathscr{M}(E\lambda)||i\rangle\right|^{2} \ .
\ee
where $\hat{\mathscr{M}}(E\lambda)\equiv\hat{H}^{(\mathbf{A}^{(\lambda)})}/(wA)$, $w\equiv [I_{f}(I_{f}+1)-I_{i}(I_{i}+1)+K_{i}^{2}]/2C_{0}$, and $k$ is the
energy (or momentum) of the photon involved in the transition.

The LO inter-band $B(E2)$ values for transitions from the $\beta$ band
to the ground band are
\begin{equation}
\label{be2beta}
B(E2,i_{\beta}\rightarrow f_{g}) = \frac{C_{\beta}^{2}}{2C_{0}^{2}\omega_{0}}\frac{q^{2}}{60}\left(C_{I_{i}020}^{I_{f}0}\right)^{2},
\end{equation}
while LO $B(E2)$ values for transitions from the $\gamma$ band to the
ground band are
\begin{equation}
\label{be2gamma}
B(E2,i_{\gamma}\rightarrow f_{g}) = \frac{3C_{\gamma}^{2}}{2C_{0}^{2}\omega_{2}}\frac{q^{2}}{60}\left(C_{I_{i}22-2}^{I_{f}0}\right)^{2}.
\end{equation}

We can generalize the definition of the quadrupole transition
moments to
\begin{equation}
Q_{if}^2=\frac{B(E2,i\rightarrow f)}{\left(C_{I_{i}K_{i}2K_{f}-K_{i}}^{I_{f}K_{f}}\right)^{2}} \ ,
\end{equation}
then
\begin{equation}
Q_{i_{\beta}f_{g}}^2=\frac{C_{\beta}^{2}}{2C_{0}^{2}\omega_{0}}Q^{2} \ , \qquad Q_{i_{\gamma}f_{g}}^2=\frac{3C_{\gamma}^{2}}{2C_{0}^{2}\omega_{2}}Q^{2},
\end{equation}
where $Q\equiv\sqrt{q^2/60}$. We note that the strengths of
transitions from the $\beta$ band are similar to those of the $\gamma$
band for similarly sized LECs $C_{\beta}$ and $C_{\gamma}$.

We note that -- within the effective theory -- the intra-band
transitions depend on the LECs $C_\beta$ and $C_\gamma$. We recall
that these LECs enter at the Hamiltonian at NNLO as off-diagonal
corrections to the Hamiltonian, which prevents us from adjusting them
to spectra at this order. As more terms enter the Hamiltonian at
N$^3$LO, it seems attractive to determine $C_\beta$ and $C_\gamma$
instead by inter-band transitions. In what follows, we adjust these
coefficients to the description of one inter-band transition from the
respective band to the ground band. Other inter-band transitions then
are predictions.

In the traditional collective models, no new parameters enter the
computation of the inter-band transitions. As a result, these faint
transitions are overpredicted substantially. For example, the
inter-band $B(E2)$ values according to the adiabatic Bohr model
are (See, e.g., Ref.~\cite{rowe2010})
\bea
B(E2,i_{\beta}\to f_{g})&=&\frac{\xi}{2\omega_{0}}\left(\frac{Ze\beta_{0}}{A}\right)^{2}\left(C_{l_{i}020}^{l_{f}0}\right)^{2} \ \nonumber\\
B(E2,i_{\gamma}\to f_{g})&=&\frac{2\xi}{\omega_{2}}\left(\frac{Ze\beta_{0}}{A}\right)^{2}\left(C_{l_{i}22-2}^{l_{f}0}\right)^{2},
\eea
implying inter-band transitions from the $\beta$ are only a
factor two weaker than those from the $\gamma$ bands. Here, $\beta_0$
is a deformation parameter. Thus, the effective theory is richer in
structure (through two additional parameters). This more complex
structure is a consequence of a theory that is based on symmetry
principles alone. It will allow us to describe inter-band transitions
much more accurately. Regarding ratios of inter-band transition
strengths, the effective theory at leading order reproduces the
traditional collective models as expected from the Alaga rules.

\subsection{Comparison with experimental data}
We test the expressions (\ref{be2beta}) and (\ref{be2gamma}) by
confronting them to data for inter-band transitions in $^{166,168}$Er
and $^{154}$Sm. These isotopes of erbium are considered good rotors,
while the samarium isotope is between rotors and transitional nuclei.

For $^{168}$Er, the relevant energies are $\xi\approx 79.8$~keV,
$\omega_{0}\approx 1217.2$~keV, and $\omega_{2}/2\approx 821.2$~keV.
In the spirit of the theory, all constants were fitted to low-energy
data. Thus, the effective quadrupole moment was fitted via the
$2_{g}^{+}\to 0_{g}^{+}$ transition, while the values
$C_{\beta}\approx 0.077$~keV$^{-1/2}$ and $C_{\gamma}\approx
0.203$~keV$^{-1/2}$ are determined from
the $2_{\beta}^{+}\to 0_{g}^{+}$ and $2_{\gamma}^{+}\to 2_{g}^{+}$
transitions, respectively. We employed data from Ref.~\cite{baglin2010} and Ref.~\cite{kotlinski1990} for completion. The values of these LECs are natural in
size when compared to the scale $\xi^{-1/2}\approx
0.112$~keV$^{-1/2}$. Clearly, more precise data for transitions
between the $\beta$ and ground bands is required to determine the size
of $C_{\beta}$. All other transitions are predictions.
Table~\ref{inter168er} shows experimental and theoretical $B(E2)$
values for transitions within the ground-state band and inter-band
transitions in $^{168}$Er. Overall, the effective theory describes the
data well. The theoretical uncertainties presented in
Table~\ref{inter168er} for transitions in the ground-state band are
based on the discussion in Subsection~\ref{uncertainties}. However,
for the uncertainties of transitions from the $\beta$ band or the
$\gamma$ band, we employed more conservative uncertainty estimates
based on the larger ratio $(\omega/\Lambda)^2\approx 0.25$ that is due
to the proximity of the breakdown scale $\Lambda$.

\begin{table}[h]
\centering
\caption{\label{inter168er} Transition strength for $^{168}$Er in
  units of e$^2$b$^2$. Experimental transitions strengths $B(E2)_{\rm
    exp}$ are compared to theoretical results $B(E2)_{\rm ET}$ from
  the effective theory and $B(E2)_{\rm BH}$ from the adiabatic Bohr
  Hamiltonian. Experimental values are taken from~\cite{baglin2010} 
  unless otherwise specified. Values for the
  adiabatic Bohr Hamiltonian are taken from~\textcite{rowe2010}. 
  Parenthesis denote experimental errors and theoretical uncertainty estimates.} 
\begin{ruledtabular}
\begin{tabular}{cD{.}{.}{9}D{.}{.}{9}D{.}{.}{4}}
$i\rightarrow f$&\multicolumn{1}{c}{$B(E2)_{\rm exp}$}&\multicolumn{1}{c}{$B(E2)_{\rm ET}$}&\multicolumn{1}{c}{$B(E2)_{\rm BH}$}\\ \hline
$2_{g}^{+} \rightarrow 0_{g}^{+}$ & 1.173 (22) & 1.173\footnotemark[2] & 1.173 \\
$4_{g}^{+} \rightarrow 2_{g}^{+}$ & 1.756 (50) & 1.676 (36) & 1.677 \\
$6_{g}^{+} \rightarrow 4_{g}^{+}$ & 2.335 (99) & 1.846 (91) & 1.842 \\
$8_{g}^{+} \rightarrow 6_{g}^{+}$ & 1.949 (72) & 1.932 (169) & 1.935 \\
\hline
$2_{\gamma}^{+} \rightarrow 0_{g}^{+}$ & 0.0258 (9) & 0.0309 (77) & 0.1126 \\
$2_{\gamma}^{+} \rightarrow 2_{g}^{+}$ & 0.0442 (38)\footnotemark[1] & 0.0442\footnotemark[2] & 0.1610 \\
$2_{\gamma}^{+} \rightarrow 4_{g}^{+}$ & 0.0034 (2) & 0.0022 (5) & 0.0080 \ \\
\hline
$2_{\beta}^{+} \rightarrow 0_{g}^{+}$ & 0.0020 (^{+8}_{-20}) & 0.0020\footnotemark[2] & 0.0387 \\
$2_{\beta}^{+} \rightarrow 2_{g}^{+}$ & & 0.0029 (7) & 0.0553 \\
$2_{\beta}^{+} \rightarrow 4_{g}^{+}$ & 0.0121 (^{+44}_{-121}) & 0.0051 (13) & 0.0995  \\
\end{tabular}
\end{ruledtabular}
\footnotetext[1]{From \textcite{kotlinski1990}.}
\footnotetext[2]{Values employed to adjust LECs of the effective theory.}
\end{table}

For $^{166}$Er, the energy scales are $\xi\approx 80.6$ keV,
$\omega_{0}\approx 1460$ keV and $\omega_{2}/2\approx 785.9$ keV. This
yields $C_{\beta}\approx 0.111$~keV$^{-1/2}$ and $C_{\gamma}\approx
0.213$~keV$^{-1/2}$, and both values are natural in size when compared
to $\xi^{-1/2}\approx 0.111$~keV$^{-1/2}$. Once again, more precise
experimental $B(E2)$ values for transitions between the $\beta$ and
ground bands would be valuable. Table~\ref{inter166er} shows
experimental and theoretical $B(E2)$ values for intra-band and
inter-band transitions in this nucleus. Theoretical uncertainties are
given as discussed for ${168}$Er.  The experimental $B(E2)$ value for
the $2_{\beta}^{+}\to 4_{g}^{+}$ transition is too large (one order of
magnitude larger than decays from the $\gamma$ band to the ground
band) to be understood within the effective theory.

\begin{table}[h]
\centering
\caption{\label{inter166er} Same as Table~\ref{inter168er} but for $^{166}$Er. Experimental values are taken from~\cite{baglin2008}.}
\begin{ruledtabular}
\begin{tabular}{c D{.}{.}{9}D{.}{.}{7}D{.}{.}{4}}
$i\rightarrow f$&\multicolumn{1}{c}{$B(E2)_{\rm exp}$}&\multicolumn{1}{c}{$B(E2)_{\rm ET}$}&\multicolumn{1}{c}{$B(E2)_{\rm BH}$}\\ \hline
\hline
$2_{g}^{+} \rightarrow 0_{g}^{+}$ & 1.175 (27) & 1.175\footnotemark[1] & 1.175 \\
$4_{g}^{+} \rightarrow 2_{g}^{+}$ & 1.718 (61) & 1.679 (24) & 1.680 \\
$6_{g}^{+} \rightarrow 4_{g}^{+}$ & 2.037 (110) & 1.849 (60) & 1.845 \\
$8_{g}^{+} \rightarrow 6_{g}^{+}$ & 2.054 (77) & 1.935 (112) & 1.939 \\
\hline
$2_{\gamma}^{+} \rightarrow 0_{g}^{+}$ & 0.0285 (12) & 0.0370 (93) & 0.1205 \\
$2_{\gamma}^{+} \rightarrow 2_{g}^{+}$ & 0.0529 (33) & 0.0529\footnotemark[1] & 0.1721 \\
$2_{\gamma}^{+} \rightarrow 4_{g}^{+}$ & 0.0043 (2) & 0.0026 (7) & 0.0086 \\
\hline
$2_{\beta}^{+} \rightarrow 0_{g}^{+}$ & 0.0036 (4)  & 0.0036\footnotemark[1] & 0.0324 \\
$2_{\beta}^{+} \rightarrow 2_{g}^{+}$ &  & 0.0051 (13) & 0.0463 \\
$2_{\beta}^{+} \rightarrow 4_{g}^{+}$ & 0.2113 (325) & 0.0093 (23) & 0.0834 \\
\end{tabular}
\end{ruledtabular}
\footnotetext[1]{Values employed to adjust the LECs of the effective theory.}
\end{table}

Let us also attempt to describe a non-rigid rotor. The region around
$^{152}$Sm has been well
studied~\cite{zamfir1999,casten2001,garrett2009}, and absolute $B(E2)$
values for some inter-band transitions in $^{154}$Sm were measured
recently~\cite{moller2012}. For $^{154}$Sm, the LECs related to
inter-band transitions are $C_{\beta}\approx 0.092$~keV$^{-1/2}$
(determined from the $2_{\beta}^{+}\to 2_{g}^{+}$ transition) and
$C_{\gamma}\approx 0.181$~keV$^{-1/2}$. Both values are natural in
size when compared to $\xi^{-1/2}\approx 0.110$~keV$^{-1/2}$.
Table~\ref{inter154sm} shows our LO results for this nucleus. The
theoretical uncertainties are computed as discussed for $^{168}$Er.
We also show theoretical results of the confined $\beta$ soft (CBS)
model~\cite{pietralla2004}, as an example that a particular model can
approximately account for the magnitude of some of the transitions
between the $\beta$ band and the ground-state band.

\begin{table}[h]
\centering
\caption{\label{inter154sm} Same as Table~\ref{inter168er} but for $^{154}$Sm. 
  Theoretical results from the confined $\beta$ soft (CBS) model~\cite{pietralla2004}, 
  taken from Ref.~\cite{moller2012}, are also included. Experimental values 
  are taken from~\cite{reich2009} and~\cite{moller2012} for intra-band 
  and inter-band transitions, respectively.}
\begin{ruledtabular}
\begin{tabular}{c D{.}{.}{9}D{.}{.}{7}D{.}{.}{4}D{.}{.}{4}}
$i\rightarrow f$&\multicolumn{1}{c}{$B(E2)_{\rm exp}$}&\multicolumn{1}{c}{$B(E2)_{\rm ET}$}&\multicolumn{1}{c}{$B(E2)_{\rm CBS}$}&\multicolumn{1}{c}{$B(E2)_{\rm BH}$}\\ \hline
\hline
$2_{g}^{+} \rightarrow 0_{g}^{+}$ & 0.863 (5) & 0.863\footnotemark[1] & 0.853 & 0.863 \\
$4_{g}^{+} \rightarrow 2_{g}^{+}$ & 1.201 (29) & 1.233 (9) & 1.231 & 1.234 \\
$6_{g}^{+} \rightarrow 4_{g}^{+}$ & 1.417 (39) & 1.358 (23) & 1.378 & 1.355 \\
$8_{g}^{+} \rightarrow 6_{g}^{+}$ & 1.564 (83) & 1.421 (43) & 1.471 & 1.424 \\
\hline
$2_{\gamma}^{+} \rightarrow 0_{g}^{+}$ & 0.0093 (10) & 0.0110 (28) & & 0.0492 \\
$2_{\gamma}^{+} \rightarrow 2_{g}^{+}$ & 0.0157 (15) & 0.0157\footnotemark[1] & & 0.0703 \\
$2_{\gamma}^{+} \rightarrow 4_{g}^{+}$ & 0.0018 (2)   & 0.0008 (2) & & 0.0050 \\
\hline
$2_{\beta}^{+} \rightarrow 0_{g}^{+}$ & 0.0016 (2)  & 0.0025 (6) & 0.0024 & 0.0319 \\
$2_{\beta}^{+} \rightarrow 2_{g}^{+}$ & 0.0035 (4) & 0.0035\footnotemark[1] & 0.0069 & 0.0456 \\
$2_{\beta}^{+} \rightarrow 4_{g}^{+}$ & 0.0065 (7) & 0.0063 (16) & 0.0348 & 0.0821 \\
\end{tabular}
\end{ruledtabular}
\footnotetext[1]{Values employed to adjust the LECs of the effective theory.}
\end{table}

We note that the ratio $C_\gamma/C_\beta$, while usually natural in
size, fulfills $C_\gamma/C_\beta > 1$ for the nuclei we just
considered. As the LECs $C_{\beta}$ and $C_\gamma$ enter quadratically
into $B(E2)$ transition strengths, the transitions from the $\beta$
band to the ground-state band are considerably weaker than the
transitions from the $\gamma$ band to the ground-state band.

The most important result of this paper is that the effective theory,
with its model-independent approach to the collective Hamiltonian and
its corresponding transition operators, suggests a step toward the
solution of the long-standing problem posed by the faint inter-band
transitions. The consistent description of Hamiltonian and currents
shows that the observed strengths of inter-band transitions can be
described within the effective theory using LECs of natural size.  As
a consequence, the strengths of the interband $E2$ transitions are
also natural in size. From this perspective it seems adequate to keep
referring to the $0_2^+$ rotational band as the $\beta$ band.  The
effective theory predicts the strength of inter-band transitions once
a single transition determines a LEC of the Hamiltonian.

Let us finally also comment on NLO corrections to inter-band
transitions. These corrections are beyond the scope of the present
paper.  Recently, \textcite{kulp2006} precisely measured {\it ratios}
of transitions intensities between the $\gamma$ band and the
ground-state band in $^{166}$Er. They confirmed the
beyond-leading-order predictions by \textcite{mikhailov1966} to a high
level of accuracy.

\section{Discussion}

The geometric collective models approach low-lying excitations in
deformed nuclei as quantized surface oscillations of a liquid drop.
In contrast, the effective theory for deformed nuclei assumes symmetry
properties (such as rotational invariance), the emergent breaking of
rotational symmetry (and the ensuing separation of scales), and the
existence of a breakdown scale. It then builds the most general
Hamiltonian (and currents) consistent with these assumptions and
orders them in magnitude based on the power counting. Not
surprisingly, the effective theory -- particularly beyond leading
order -- has more parameters than the traditional models.  The
geometrical models quantitatively predict several aspects of deformed
nuclei, e.g. rotational bands with similar-sized rotational constants
on top of intrinsic vibrations together with strong in-band
transitions.  The effective theory obtains these results in leading
and subleading order.

Other aspects, such as the small variation of rotational constants
with the quantum numbers of the band heads, or the magnitude of
inter-band transitions are not described quantitatively correct by the
tradtional models. In contrast, the effective theory also captures
these finer details, as shown for the rotational constants in
Ref.~\cite{zhang2013} and for the inter-band transitions in this work.
This suggests that the assumptions made by the models are correct only
to a certain order. The effective theory's capability in accounting
also for the finer details suggest that its underlying assumptions are
sound.  The effective theory delivers increased precision (with
consistent uncertainty estimates) at the expense of additional
parameters. This can be useful if correspondingly precise data is
available, which is the case for in-band transitions in transitional
nuclei and for inter-band transitions considered in this work. This
aspect is also of interest with view on the advent of powerful
$\gamma$-ray detectors~\cite{paschalis2013}.

The capability to estimate theoretical uncertainties is essential when
confronting theory and experiment. It is natural to effective theories
because of their power counting. In addition, the identification of a
breakdown scale makes clear up to which energies the theory can be
applied. We believe Figs.~\ref{rotbe2} and \ref{rotbe2sm} would carry
little information without the theoretical uncertainty
estimates. These estimates motivate us to propose re-measurements or
re-evaluations of certain data.

\section{Summary}

We studied $E2$ transitions in deformed nuclei within a
model-independent approach based on an effective theory. The effective
theory is based on the emergent symmetry breaking of rotational
symmetry to axial symmetry.  Electromagnetic transitions result from
gauging of the Hamiltonian, and from higher-order non-minimal
couplings that are consistent with gauge invariance and the symmetry
of the system under consideration. The estimate of theoretical
uncertainties is one of the highlights of the effective theory
approach.

Homonuclear molecules provide us with an ideal test case because they
possess a very large separation of scale and therefore exhibit only
small corrections to the rigid-rotor limit. The effective theory
describes $E2$ transitions in the diatomic molecules N$_2$ and H$_2$
very well, and deviations are within the theoretical uncertainties.

The effective theory describes $B(E2)$ transitions in the ground-state
bands of well-deformed nuclei at leading order, and more precise
experimental data are necessary to probe subleading effects. Our
model-independent results also suggest that some low-lying transitions
in these nuclei would probably merit a more precise re-measurement or
re-evaluation of data, because they can not be easily understood
within the effective theory.  For transitional nuclei, the existing
data are sufficiently precise to probe the effective theory at
subleading oder. Here, data and theoretical results are
consistent within theoretical uncertainties.

For $E2$ transitions within ground-state bands, the effective theory
reproduces known results of the Bohr Hamiltonian. The employment of
the beakdown scale and the power counting allows us to estimate
theoretical uncertainties and to meaningfully confront data. A somewhat
surprising result is that well-deformed nuclei do not challenge theory
because of insufficient precision of the available data.

The effective theory also suggests that the electromagnetic structure
of deformed nuclei is more complex than the collective models assume,
regarding both the Hamiltonian and the transition operator.  The
magnitude of the faint inter-band transitions is captured correctly
within the effective theory, thus addressing to a long-standing
problem. In the effective theory, this comes at the expense of new
parameters, and one needs to know a single inter-band transition
strength to make leading-order predictions for other transitions
between the bands in question. These results also cast some doubt on
the traditional usage of the quadrupole operator to describe faint
electromagnetic transitions, as this approach seems to be limited to
the strong (leading order) transitions between states within a band. 

This work shows that the effective theory for deformed nuclei
reproduces the traditional collective models regarding leading-order
aspects (spectra and transitions) of deformed nuclei. In contrast to
the models, however, the effective theory also accounts for finer
details, and it provides us with theoretical error estimates.  
We would hope that the results presented in this work
might stimulate more precise measurements of electromagnetic
transitions in deformed nuclei.

\begin{acknowledgments}
  We thank M. Allmond, M. Caprio, A. Ekstr{\"o}m, C. Forss{\'e}n,
  R.~J. Furnstahl, H. Griesshammer, H.-W. Hammer, K. Jones, H.~Krebs,
  and L.~Platter for useful discussions. This material is based upon
  work supported by the U.S.  Department of Energy, Office of Science,
  Office of Nuclear Physics under Award Number DEFG02-96ER40963
  (University of Tennessee), and under Contract No.  DE-AC05-00OR22725
  (Oak Ridge National Laboratory).
\end{acknowledgments}


\end{document}